\documentclass[aps,prd,nofootinbib,showpacs,superscriptaddress,twocolumn,floatfix]{revtex4-1}

\usepackage[utf8]{inputenc}
\usepackage{mathrsfs}
\usepackage{bm}
\usepackage{bbm}
\usepackage{indentfirst}
\usepackage{epstopdf}
\usepackage{color}
\usepackage{xcolor}
\usepackage{amssymb, amsmath,color, hyperref, graphicx}
\usepackage{braket}
\usepackage{placeins}
\usepackage{multirow}
\usepackage{slashed}
\usepackage{physics}
\usepackage{booktabs}
\usepackage{tikz}
\usepackage[mathlines]{lineno}
\usepackage{comment}
\usetikzlibrary{arrows.meta}

\begin{document}

\title{Static Quark--Antiquark Interactions in Rotating SU(3) Gluodynamics}

\author{Heng-Tong Ding}
\address{Key Laboratory of Quark and Lepton Physics (MOE) and Institute of
Particle Physics, Central China Normal University, Wuhan 430079, China}
\author{Olaf Kaczmarek}
\affiliation{Fakultät für Physik, Universität Bielefeld, D-33615 Bielefeld, Germany}
\author{Ran Luo}
\affiliation{Key Laboratory of Quark and Lepton Physics (MOE) and Institute of
Particle Physics, Central China Normal University, Wuhan 430079, China}
\author{Hai-Tao Shu}
\affiliation{Key Laboratory of Quark and Lepton Physics (MOE) and Institute of
Particle Physics, Central China Normal University, Wuhan 430079, China}

\begin{abstract}
We study static quark--antiquark interactions in rotating SU(3) gluodynamics using quenched lattice simulations at imaginary angular velocity. At zero temperature, we extract the static potential from Wilson loops for quark--antiquark pairs aligned with the rotation axis, for transverse pairs with one source on the rotation axis, and for symmetric transverse pairs across the rotation axis. Within the present accuracy, no significant rotation dependence or anisotropy is observed in the zero-temperature potential.
At finite temperature, imaginary rotation suppresses the color-averaged free energies obtained from Polyakov-loop correlators in both longitudinal and transverse geometries. Axial–diagonal comparisons are used to identify a bulk region where open-boundary artifacts are reduced. In this region, the large-distance longitudinal free-energy shift is well described by $\Delta F_z(R_{xy})=A R_{xy}^2+B$. The transverse channels exhibit the same qualitative suppression, while their distance dependence additionally reflects the radial arrangement of the static sources and is compatible with a radial single-source free-energy shift in the bulk region. For the finite-temperature observables studied above $T_c$, the response weakens as the temperature is increased.
These results provide lattice evidence for a position- and geometry-dependent response of bare static-source free energies to imaginary rotation in a gluonic medium.

\end{abstract}

\maketitle


\section{Introduction}
\label{sec:introduction}

Non-central relativistic heavy-ion collisions can generate a large orbital angular momentum and create strongly interacting matter with sizable vorticity. 
The observation of global polarization of $\Lambda$ and $\bar{\Lambda}$ hyperons provides experimental evidence that spin degrees of freedom are sensitive to the vortical structure of the produced quark--gluon plasma~\cite{STAR:2017ckg,STAR:2021beb,Becattini:2020ngo}. 
This has motivated broad theoretical interest in QCD matter under rotation, including anomalous transport phenomena such as the chiral vortical effect, possible modifications of the QCD phase structure, and changes of hadronic and heavy-quark observables in a rotating medium~\cite{Kharzeev:2015znc,Fukushima:2018grm}.

A first-principles treatment of QCD in a rotating frame was formulated by Yamamoto and Hirono~\cite{Yamamoto:2013zwa}. 
In this approach, rotation is introduced through the non-inertial metric of a co-rotating frame, or equivalently through an external gravitational background. 
In Euclidean simulations, real angular velocity leads to a sign problem, and lattice calculations are therefore usually performed at imaginary angular velocity $\Omega_I$, followed by an analysis of the dependence on $\Omega_I^2$ and, when applicable, analytic continuation to real angular velocity. 
This formulation has been used in recent lattice studies of rotating gluonic matter and QCD thermodynamics~\cite{Braguta:2020biu,Braguta:2021jgn,Yang:2023vsw,Braguta:2022str,Braguta:2025yud,Chernodub:2022veq,Braguta:2023yjn,Braguta:2023iyx,Braguta:2024zpi}.

In SU(3) gluodynamics, lattice studies of the confinement--deconfinement transition based on the Polyakov loop and its susceptibility found a nontrivial dependence of the transition temperature on rotation~\cite{Braguta:2020biu,Braguta:2021jgn}. 
Using simulations in the rotating-frame setup and analyzing the angular-velocity dependence, these studies concluded that the critical temperature of the confinement--deconfinement transition in gluodynamics increases with increasing real angular velocity. 
They also emphasized that boundary conditions in the plane perpendicular to the rotation axis play an important role in finite-volume simulations of rotating systems. 
The extension to QCD under rotation with dynamical quarks has also been explored. 
For $N_f=2$ clover-improved Wilson quarks, separate rotations of the gluonic and fermionic sectors were found to have competing effects on the pseudocritical temperatures, while the combined effect leads to an increase in the pseudocritical temperatures with angular velocity~\cite{Braguta:2022str}. 
For $N_f=2+1$ staggered fermions, finite-temperature QCD has been studied at imaginary angular velocity; after analytic continuation, the results were interpreted as rotational catalysis of chiral symmetry breaking and confinement, qualitatively consistent with the available $N_f=2$ Wilson-fermion study~\cite{Yang:2023vsw,Braguta:2022str}. 
These studies show that lattice QCD can provide direct information on the thermodynamics of rotating strongly interacting matter, while also emphasizing that spatially resolved observables and correlation functions require special care because rotation introduces a preferred axis and a radial dependence.

Recent lattice studies have further shown that rotating gluonic matter can be spatially inhomogeneous. 
At imaginary angular velocity, a radially dependent Polyakov loop has been observed in rapidly rotating SU(3) Yang--Mills plasma and discussed in connection with the Tolman--Ehrenfest relation in imaginary time~\cite{Chernodub:2022veq}. 
Other simulations of rotating gluodynamics reported unusual mechanical properties, including a negative moment of inertia in certain temperature ranges~\cite{Braguta:2023yjn}, and identified a mixed inhomogeneous phase with spatially separated confining and deconfining regions~\cite{Braguta:2023iyx,Braguta:2024zpi}. 
These results indicate that, in a rotating system, local observables can depend strongly on the distance from the rotation axis. 
They also show that boundary conditions in the transverse plane are not merely a technical detail, but an essential ingredient in defining and interpreting lattice observables in a finite rotating volume.

Theoretical studies based on effective models, perturbation theory, holography, and strong-coupling expansions have led to a diverse picture. 
In NJL-type, quark-meson, and Polyakov-loop-extended models, rotation can affect the chiral and deconfinement sectors differently, often favoring chiral symmetry restoration while producing a model-dependent response of the Polyakov-loop sector~\cite{Jiang:2016wvv,Chernodub:2016kxh,Wang:2018sur,Chen:2023cjt,Sun:2023kuu,Cao:2023olg,Sun:2024anu}. 
At imaginary angular velocity, perturbative and strong-coupling studies find confinement-favoring effects, including perturbative or spatially inhomogeneous confinement and an increasing deconfinement temperature with increasing $\Omega_I$~\cite{Chen:2022smf,Chen:2024tkr,Fukushima:2025hmh}. 
For real angular velocity, strong-coupling and holographic studies also find sizable rotational effects, but the resulting trends in the transition temperature and string tension depend on the formulation~\cite{Wang:2025mmv,Chen:2024edy}. 
These results indicate that rotating QCD thermodynamics is sensitive to the treatment of imaginary versus real angular velocity, analytic continuation, and the definition of the rotating ensemble.

Static quark--antiquark interactions provide a direct probe of confinement and color screening. 
At zero temperature, the static potential extracted from Wilson loops characterizes the interaction between infinitely heavy color sources. 
At finite temperature, Polyakov-loop correlators determine color-averaged free energies and encode the screening of static color charges in the medium~\cite{Kaczmarek:2005ui,Burnier:2017bod}. 
In a rotating system, these observables can depend not only on the quark--antiquark separation, but also on the orientation of the pair relative to the rotation axis and on the radial positions of the static sources. 
They are therefore natural probes of possible orientation- and position-dependent responses of static-source observables in a rotating gluonic medium. 
Related questions have been studied in holographic models, where rotation was found to modify the heavy-quark potential, drag force, string breaking, running coupling, screening distance, string tension, and Polyakov loop in an orientation- and position-dependent manner~\cite{Chen:2023yug,Zhou:2023qtr,Chen:2024edy}. 
A direct lattice study of static quark--antiquark potentials and free energies in rotating SU(3) gluodynamics, however, has not yet been reported.

In this work we study static quark--antiquark interactions in rotating SU(3) gluodynamics using quenched lattice simulations at imaginary angular velocity. 
At zero temperature, we extract static potentials from Wilson loops for three geometries relative to the rotation axis: quark--antiquark pairs separated along the rotation axis, transverse pairs with one source located on the rotation axis, and symmetric transverse pairs placed across the rotation axis. 
At finite temperature, we determine the corresponding color-averaged free energies from Polyakov-loop correlators. 
Since open boundaries in the transverse plane break translational invariance and induce boundary artifacts, we compare axial and diagonal directions in the transverse plane to identify a bulk region where such effects are under control. 
This allows us to quantify the dependence of the rotation-induced free-energy shifts on angular velocity, temperature, and distance from the rotation axis.

The paper is organized as follows. 
Section~\ref{sec:framework} defines the Wilson-loop potentials and Polyakov-loop free energies used in this study. 
Section~\ref{sec:Lattice-setup} describes the rotating-frame lattice action, boundary conditions, simulation parameters, and scale setting. 
Sections~\ref{sec:results_zeroT} and \ref{sec:results_finiteT} present the zero- and finite-temperature results, including the analysis of boundary effects and the temperature dependence of the rotation-induced free-energy shifts. 
Section~\ref{sec:summary} summarizes our conclusions and discusses possible extensions.

\section{Static quark–antiquark interactions under rotation}
\label{sec:framework}
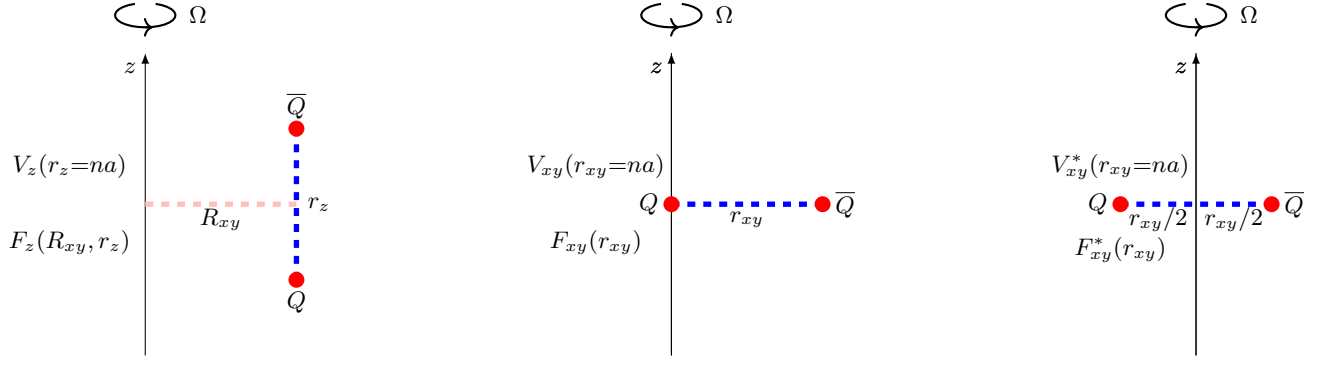
\begin{figure*}[!htp]
	\centerline{
		\begin{tikzpicture}
		\draw[-latex] (0,-2)--(0,2);\node at(-0.2,1.8){$z$};\draw [thick,domain=110:430] plot ({0.4*cos(\x)},{2.5+0.16*sin(\x)});\draw[->,thick](0,2.34)--(0.1,2.34);\node at(0.7,2.5){$\Omega$};
		\draw[pink,line width=2pt,dashed] (0,0)--(2,0);\node at(1,-0.2){$R_{xy}$};
		\draw[blue,line width=2pt,dashed] (2,-1)--(2,1);\node at(2.3,0){$r_z$};
		\fill[red] (2,-1)circle(3pt) (2,1)circle(3pt);\node at(2,-1.3){$Q$};\node at(2,1.3){$\overline{Q}$};
		\node at(-1,0.5){$V_z(r_z$=$na)$};\node at(-1,-0.5){$F_z(R_{xy},r_z)$};
		\end{tikzpicture}
		\quad\quad\quad\quad\quad\quad\quad
		\begin{tikzpicture}
		\draw[-latex] (0,-2)--(0,2);\node at(-0.2,1.8){$z$};\draw [thick,domain=110:430] plot ({0.4*cos(\x)},{2.5+0.16*sin(\x)});\draw[->,thick](0,2.34)--(0.1,2.34);\node at(0.7,2.5){$\Omega$};
		\draw[black] (0,-2)--(0,2);\node at(-0.2,1.8){$z$};
		\draw[blue,line width=2pt,dashed] (0,0)--(2,0);\node at(1,-0.2){$r_{xy}$};
		\fill[red] (2,0)circle(3pt) (0,0)circle(3pt);\node at(-0.3,0){$Q$};\node at(2.3,0){$\overline{Q}$};
		\node at(-1,0.5){$V_{xy}(r_{xy}$=$na)$};\node at(-1,-0.5){$F_{xy}(r_{xy})$};
		\end{tikzpicture}
		\quad\quad\quad\quad\quad\quad\quad
		\begin{tikzpicture}
		\draw[-latex] (0,-2)--(0,2);\node at(-0.2,1.8){$z$};\draw [thick,domain=110:430] plot ({0.4*cos(\x)},{2.5+0.16*sin(\x)});\draw[->,thick](0,2.34)--(0.1,2.34);\node at(0.7,2.5){$\Omega$};
		\draw[black] (0,-2)--(0,2);\node at(-0.2,1.8){$z$};
		\draw[blue,line width=2pt,dashed] (-1,0)--(1,0);\node at(0.5,-0.2){$r_{xy}/2$};\node at(-0.5,-0.2){$r_{xy}/2$};
		\fill[red] (-1,0)circle(3pt) (1,0)circle(3pt);\node at(-1.3,0){$Q$};\node at(1.3,0){$\overline{Q}$};
		\node at(-1,0.5){$V_{xy}^*(r_{xy}$=$na)$};\node at(-1,-0.6){$F_{xy}^*(r_{xy})$};
		\end{tikzpicture}
	}
	\caption{Schematic illustration of the three static-source geometries considered
in this work.  The quantities $V_z$, $V_{xy}$, and $V^*_{xy}$ denote
the static potentials extracted from Wilson loops at $T\simeq0$, while
$F_z$, $F_{xy}$, and $F^*_{xy}$ denote the corresponding color-averaged
free energies extracted from Polyakov-loop correlators at finite
temperature.}
	\label{fig:potential}
\end{figure*}

In this section we define the Wilson-loop potentials and
Polyakov-loop free energies used to study static quark--antiquark
interactions in the rotating system. We specifically explore three distinct spatial configurations of heavy quark pairs relative to the rotation axis ($z$-axis), as illustrated schematically in~\autoref{fig:potential}.

\subsection{Zero Temperature Potentials}
\label{subsec:zeroT_conf}
At zero temperature, we utilize Wilson loops to determine the static
potential between a quark--antiquark pair. For a given geometry $s$,
we define
\begin{equation}
\label{eq:Wilson_loop_def}
W_s(n,n_\tau)=
\mathrm{Tr}\left[
\prod_{(\ell,\mu)\in\mathcal{L}_s(n,n_\tau)}
U_\mu(\ell)
\right],
\qquad s=z,xy,xy^* .
\end{equation}
Here $\mathcal{L}_s(n,n_\tau)$ denotes the closed rectangular contour
with spatial extent $n$ and temporal extent $n_\tau$ for the geometry
$s$, while $\ell$ labels the lattice sites along the contour and $\mu$
the corresponding link directions. The explicit contours used for the
three geometries are given in Eqs.~\eqref{eq:Wz}--\eqref{eq:Wxy_c}.

The static heavy-quark potentials are extracted from the large-$n_\tau$
limit of the logarithmic ratio,
\begin{equation}
\label{eq:static_potential_logratio}
aV_s(na)=
\lim_{n_\tau\to \infty}
\ln
\frac{\langle W_s(n,n_\tau)\rangle}
     {\langle W_s(n,n_\tau+1)\rangle},
\qquad s=z,xy,xy^* .
\end{equation}

Due to anisotropy introduced by rotation, we define the Wilson loops separately for three specific configurations:
\paragraph{Quarks aligned along the rotation axis ($z$-direction):} The loop $W_z(n,n_{\tau})$ follows the rectangular path
\begin{equation}
\label{eq:Wz}
\begin{split}
\mathcal{L}_z:
(x,y,z,\tau)\rightarrow(x,y,z+n,\tau)&\rightarrow(x,y,z+n,\tau+n_{\tau})\\
\rightarrow(x,y,z,\tau+n_{\tau})&\rightarrow(x,y,z,\tau)\,.
\end{split}
\end{equation}
For the $T\simeq 0$ Wilson-loop analysis, we average this loop over
the coordinates $x$, $y$, $z$ and $\tau$ to improve statistics.  Possible
residual dependence on the radial distance $R_{xy}$ is checked
a posteriori through the comparison with the transverse geometries
shown below. This is illustrated in the left panel of~\autoref{fig:potential}.

\paragraph{One quark on the rotation axis and one off-axis in the $xy$-plane:} The Wilson loop $W_{xy}(n,n_{\tau})$ follows the loop path 
\begin{equation}
\label{eq:Wxy}
\begin{split}
\mathcal{L}_{xy}:
(0,0,z,\tau)\rightarrow\left\{
\begin{aligned}
(\pm n,0,z,\tau)\\
(0,\pm n,z,\tau)
\end{aligned}
\right.
\rightarrow&\left\{
\begin{aligned}
(\pm n,0,z,\tau+n_{\tau})\\
(0,\pm n,z,\tau+n_{\tau})
\end{aligned}
\right.\\
\rightarrow(0,0,z,\tau+n_{\tau})\rightarrow&(0,0,z,\tau),
\end{split}
\end{equation}
averaged over the temporal and longitudinal spatial coordinates ($\tau$ and $z$). This is illustrated in the middle panel of~\autoref{fig:potential}.

\paragraph{Symmetric quark--antiquark pair across the rotation axis in the $xy$-plane:}
The symmetric transverse Wilson loop $W_{xy}^*(n=2k,n_\tau)$ is
constructed from the two contours $\mathcal{L}_x^*$ and
$\mathcal{L}_y^*$,
\begin{equation}
\label{eq:Wxy_c}
\begin{split}
\mathcal{L}_x^*:~
&(-k,0,z,\tau)\rightarrow(k,0,z,\tau)
\rightarrow(k,0,z,\tau+n_\tau)\\
&\rightarrow(-k,0,z,\tau+n_\tau)
\rightarrow(-k,0,z,\tau),\\
\mathcal{L}_y^*:~
&(0,-k,z,\tau)\rightarrow(0,k,z,\tau)
\rightarrow(0,k,z,\tau+n_\tau)\\
&\rightarrow(0,-k,z,\tau+n_\tau)
\rightarrow(0,-k,z,\tau).
\end{split}
\end{equation}
The corresponding Wilson loops are averaged over the two transverse
orientations, as well as over $z$ and $\tau$; see the right panel
of~\autoref{fig:potential}.

\subsection{Finite Temperature Free Energies}
\label{subsec:finiteT_conf}
At finite temperature, we employ Polyakov loop correlators to calculate color-averaged free energies. We define the Polyakov loop as 
\begin{equation} 
\tilde{L}(R_{xy},z)=\frac{1}{3}\mathrm{Tr}\prod_{\tau=0}^{N_\tau-1}U_4(x,y,z,\tau), \quad R_{xy}=\sqrt{x^2+y^2}. 
\label{eq:PolyakovLoop}
\end{equation}

Analogously, we consider the same three spatial configurations as at zero temperature, as illustrated in the left, middle and right panels of~\autoref{fig:potential}, respectively:

\paragraph{Along the rotation axis ($z$-direction):} 
\begin{equation} 
F_z(R_{xy},r_z)=-T \ln\left\langle\frac{1}{N_z}\sum_{z=0}^{N_z-1}\tilde{L}(R_{xy},z)\tilde{L}^{\dagger}(R_{xy},z+r_z)\right\rangle. 
\label{eq:Fz_def}
\end{equation}
This quantity describes the free energy of a quark--antiquark pair separated along the rotation axis ($z$-direction), with a fixed radial distance $R_{xy}$ from the rotation axis. Here, $r_z$ denotes the separation distance along the rotation axis, and averaging is performed over spatial slices in the $z$-direction.

\paragraph{One quark on-axis, one off-axis in the $xy$-plane:} 
\begin{equation} 
F_{xy}(r_{xy})=-T \ln\left\langle\frac{1}{N_z}\sum_{z=0}^{N_z-1}\tilde{L}(0,z)\tilde{L}^{\dagger}(r_{xy},z)\right\rangle. 
\label{eq:Fxy_def}
\end{equation}
This expression represents the free energy of a quark--antiquark pair positioned in the transverse plane ($xy$-plane), with one quark fixed at the rotation axis and the other at a distance $r_{xy}$ away. The averaging is taken along the longitudinal direction ($z$), effectively isolating transverse-plane rotational effects.

\paragraph{Symmetric quark--antiquark pair across the rotation axis in the $xy$-plane:} 
\begin{equation} 
F_{xy}^*(r_{xy})=-T \ln\left\langle\frac{1}{N_z}\sum_{z=0}^{N_z-1}\tilde{L}(\vec{r}_{xy}/2,z)\tilde{L}^{\dagger}(-\vec{r}_{xy}/2,z)\right\rangle. 
\label{eq:Fxy_star_def}
\end{equation}
This definition describes the free energy between a symmetric quark--antiquark pair in the transverse ($xy$) plane, positioned symmetrically about the rotation axis, separated by a distance $r_{xy}$. This symmetric geometry places the two static sources at the same
distance from the rotation axis and is therefore useful for separating,
at least qualitatively, radial-position effects from separation effects.

To isolate rotation effects, we define the free energy difference 
\begin{equation} 
\Delta F_{s}(\cdots) =F_s(\cdots)\big|_{\Omega_I=0} - F_s(\cdots)\big|_{\Omega_I\neq 0} , \quad s=z, xy, xy^*.
\label{eq:DeltaF_def}
\end{equation}
With this convention, a positive value of $\Delta F_s$ indicates that
imaginary rotation lowers the corresponding free energy.
We use these
differences below to characterize the rotation-induced response at fixed
temperature and geometry.

To determine $T_c$ at $\Omega_I=0$, we calculate the Polyakov loop
\begin{equation}
L=\frac{1}{(N_{\sigma}+1)^2N_{\sigma}}{\rm{Tr}}\sum_{\vec{x}}\prod_{\tau=0}^{N_{\tau}-1}U_4(\vec{x},\tau)\,,
\label{eq:L_def}
\end{equation}
and its susceptibility
\begin{equation}
\chi=(N_{\sigma}+1)^2N_{\sigma}(\langle|L|^2\rangle-\langle|L|\rangle^2)\,.
\label{eq:chi_def}
\end{equation}
The critical temperature $T_c$ is determined from the peak position of the Polyakov-loop susceptibility, obtained by fitting the data near the maximum with a Gaussian ansatz. In this paper, $T_c$ always refers to the critical temperature at $\Omega_I=0$.

\section{Lattice setup}
\label{sec:Lattice-setup}
We use the framework formulated in \cite{Yamamoto:2013zwa}. In the non-inertial frame rotating with the system, SU(3) gluodynamics
can be formulated in a curved spacetime
(or, equivalently, in an external gravitational background) described by the following Euclidean metric for rotation about the $z$ axis,

\begin{equation}
g_{\mu\nu}=
\begin{pmatrix}
    1     &      0     &      0      &    y\Omega_I   \\
    0     &      1     &      0      &   -x\Omega_I   \\
    0     &      0     &      1      &       0        \\
y\Omega_I & -x\Omega_I &      0      & 1+r^2\Omega_I^2
\end{pmatrix},
\end{equation}
where $r=\sqrt{x^2+y^2}$ is the distance from the rotation axis. We have used an imaginary angular velocity $\Omega_I=-i\Omega$ to avoid the sign problem. Note that the spatial submatrix of this metric is $\mathbb{I}_{3\times 3}$, so distances such as $r_z$ and $r_{xy}$ are measured as usual. With this metric, the standard Wilson action becomes
\begin{equation}
\begin{split}
    S_G=&\beta\sum_x\Big\{
    (1+r^2\Omega_I^2) (1-\frac{1}{N_c}{\rm{ReTr}}\overline{U}_{xy})\\
    &+(1+y^2\Omega_I^2) (1-\frac{1}{N_c}{\rm{ReTr}}\overline{U}_{xz})\\
    &+(1+x^2\Omega_I^2) (1-\frac{1}{N_c}{\rm{ReTr}}\overline{U}_{yz})\\
    &+3 - \frac{1}{N_c}{\rm{ReTr}}(\overline{U}_{x\tau}+\overline{U}_{y\tau}+\overline{U}_{z\tau})\\
    &-\frac{1}{N_c}{\rm{ReTr}}\big[ y\Omega_I(\overline{V}_{xy\tau}+\overline{V}_{xz\tau}) -\\
    &x\Omega_I(\overline{V}_{yx\tau}+\overline{V}_{yz\tau}) + xy\Omega_I^2\overline{V}_{xzy} \big]
    \Big\},
\end{split}
\end{equation}
The clover-type average of
plaquettes is defined as 
\begin{equation}
    \overline{U}_{\mu\nu}=\frac{1}{4}\Bigg(
    \begin{tikzpicture}[baseline={([yshift=1pt]current bounding box.center)}]
    \draw[->,dashed](-1.2,0)--(1.2,0);\node at(1.1,0.2) {$\mu$};
    \draw[->,dashed](0,-1.2)--(0,1.2);\node at(-0.2,1.1) {$\nu$};
    \draw[red,line width=1.5pt] (-0.9,0.1)--(-0.1,0.1)--(-0.1,0.9)--(-0.9,0.9)--(-0.9,0.1);
    \draw[red,line width=1.5pt] (0.9,0.1)--(0.1,0.1)--(0.1,0.9)--(0.9,0.9)--(0.9,0.1);
    \draw[red,line width=1.5pt] (-0.9,-0.1)--(-0.1,-0.1)--(-0.1,-0.9)--(-0.9,-0.9)--(-0.9,-0.1);
    \draw[red,line width=1.5pt] (0.9,-0.1)--(0.1,-0.1)--(0.1,-0.9)--(0.9,-0.9)--(0.9,-0.1);
    \end{tikzpicture}
    \Bigg).
\end{equation}
Similarly, the chair-type average entering the action is defined as
\begin{equation}
    \overline{V}_{\mu\nu\rho}=\frac{1}{8}\Bigg(
    \begin{tikzpicture}[baseline={([yshift=1pt]current bounding box.center)}]
    \draw[->,dashed](-1.2,0,0)--(1.2,0,0);\node at(1.2,0,0.4){$\mu$};
    \draw[->,dashed](0,-1.2,0)--(0,1.2,0);\node at(0,1.2,0.4){$\rho$};
    \draw[->,dashed](0,0,1.6)--(0,0,-1.6);\node at(0.2,0,-1.4){$\nu$};
    \draw[blue,line width=1.pt] (-0.05,-0.05,-0.1)--(-0.95,-0.05,-0.1)--(-0.95,-0.05,-0.9)--(-0.05,-0.05,-0.9)--(-0.05,-0.95,-0.9)--(-0.05,-0.95,-0.1)--(-0.05,-0.05,-0.1);
    \draw[blue,line width=1.pt] (-0.05,-0.05,0.1)--(-0.95,-0.05,0.1)--(-0.95,-0.05,0.9)--(-0.05,-0.05,0.9)--(-0.05,-0.95,0.9)--(-0.05,-0.95,0.1)--(-0.05,-0.05,0.1);
    \draw[blue,line width=1.pt] (0.05,0.05,-0.1)--(0.95,0.05,-0.1)--(0.95,0.05,-0.9)--(0.05,0.05,-0.9)--(0.05,0.95,-0.9)--(0.05,0.95,-0.1)--(0.05,0.05,-0.1);
    \draw[blue,line width=1.pt] (0.05,0.05,0.1)--(0.95,0.05,0.1)--(0.95,0.05,0.9)--(0.05,0.05,0.9)--(0.05,0.95,0.9)--(0.05,0.95,0.1)--(0.05,0.05,0.1);
    \end{tikzpicture}
    -
    \begin{tikzpicture}[baseline={([yshift=1pt]current bounding box.center)}]
    \draw[->,dashed](-1.2,0,0)--(1.2,0,0);\node at(1.2,0,0.4){$\mu$};
    \draw[->,dashed](0,-1.2,0)--(0,1.2,0);\node at(0,1.2,0.4){$\rho$};
    \draw[->,dashed](0,0,1.6)--(0,0,-1.6);\node at(0.2,0,-1.4){$\nu$};
    \draw[blue,line width=1.pt] (0.05,-0.05,-0.1)--(0.95,-0.05,-0.1)--(0.95,-0.05,-0.9)--(0.05,-0.05,-0.9)--(0.05,-0.95,-0.9)--(0.05,-0.95,-0.1)--(0.05,-0.05,-0.1);
    \draw[blue,line width=1.pt] (0.05,-0.05,0.1)--(0.95,-0.05,0.1)--(0.95,-0.05,0.9)--(0.05,-0.05,0.9)--(0.05,-0.95,0.9)--(0.05,-0.95,0.1)--(0.05,-0.05,0.1);
    \draw[blue,line width=1.pt] (-0.05,0.05,-0.1)--(-0.95,0.05,-0.1)--(-0.95,0.05,-0.9)--(-0.05,0.05,-0.9)--(-0.05,0.95,-0.9)--(-0.05,0.95,-0.1)--(-0.05,0.05,-0.1);
    \draw[blue,line width=1.pt] (-0.05,0.05,0.1)--(-0.95,0.05,0.1)--(-0.95,0.05,0.9)--(-0.05,0.05,0.9)--(-0.05,0.95,0.9)--(-0.05,0.95,0.1)--(-0.05,0.05,0.1);
    \end{tikzpicture}
    \Bigg).
\end{equation}

We impose periodic boundary conditions (PBC) in the Euclidean time $\tau$ and along the rotation axis $z$. In the transverse $x$--$y$ plane, periodic boundary conditions are not compatible with the velocity profile of a rigidly rotating system and are therefore not appropriate for the rotating-frame setup. We have examined the two transverse boundary prescriptions commonly used in rotating lattice studies, namely Dirichlet and open boundary conditions.
Dirichlet boundary conditions amount to fixing all boundary links to the unit matrix, $U_\mu(x)=\mathbbm{1}$. This explicitly breaks the $\mathbb{Z}_3$ center symmetry and pins the Polyakov loop to $\mathrm{Tr}\,\mathbbm{1}=3$ at the boundary. 
For the Wilson loops and Polyakov-loop correlators considered here, such fixed boundary links lead to large unphysical contributions when the loops or correlators approach the boundary. We find this effect to be particularly severe at $T\simeq 0$. Therefore, Dirichlet boundary conditions are used only as a diagnostic check, while all results presented below are obtained with open boundary conditions (OBC) in the transverse directions.

The OBC prescription used in this work is implemented directly in the gauge action. We include only elementary plaquettes and chair loops that are fully contained inside the lattice volume; plaquettes or chair loops that would extend beyond the transverse boundary are omitted. Boundary links are
not fixed, in contrast to the Dirichlet case, but are dynamical links
with fewer action terms attached to them.
Following Ref.~\cite{Braguta:2021jgn}, this construction can be interpreted as attaching the simulated volume to a classical zero-temperature Yang–Mills background outside the lattice boundary. In the present work, however, its practical role is to reduce boundary artifacts in static-source observables.

Our simulations are carried out in the quenched approximation on $N_x\times N_y\times N_z\times N_{\tau}=(N_{\sigma}+1)\times (N_{\sigma}+1)\times N_{\sigma}\times N_{\tau}$ lattices at several values of the gauge coupling $\beta$. We take $N_\sigma=64$ throughout. For the low-temperature ensembles we use $N_\tau=64$, while for the
finite-temperature ensembles we use $N_\tau=8$. The extra lattice slice in each of the transverse directions, i.e. $N_x=N_y=N_\sigma+1$, accounts for the open boundaries imposed in $x$ and $y$. The rotation axis is chosen to pass through the geometric center of the lattice and to be
perpendicular to the $x$--$y$ plane. We label the transverse coordinates as
$x,y\in\{-R,\ldots,0,\ldots,+R\}$ with $R=N_\sigma/2$. The simulation parameters and ensemble statistics are summarized in~\autoref{table:nconf}.

For the low-temperature ensembles, we fix the gauge coupling to $\beta=6.5$, corresponding to a lattice spacing
$a=0.042~\mathrm{fm}$. For $N_\tau=64$ this gives 
$T=1/(aN_\tau)=73.5~\mathrm{MeV}$~\footnote{Since this temperature is far below the deconfinement temperature, we refer to these ensembles as the $T\simeq0$ ensembles in what follows.}. The finite-temperature ensembles are generated at the values of $\beta$ listed in~\autoref{table:nconf}.
The parameter $\Omega_I$ controls the (imaginary) angular velocity and sets the
maximum linear velocity at the corners of the $x$--$y$ plane.
At zero temperature, $\Omega_I=40~\mathrm{MeV}$ corresponds to
$v_{\max}\simeq 0.385\,c$, where $c$ denotes the speed of light. At finite temperature, due to the larger lattice spacing, a set of slightly smaller $\Omega_I$ are used. The maximum linear velocity occurs at $\beta=6.1$, $\Omega_I=30~\mathrm{MeV}$, which is $0.512\,c$.

For scale setting we employ the parametrization proposed in Ref.~\cite{Francis:2015lha},
using the updated fit parameters given in Ref.~\cite{Burnier:2017bod}.

Statistical uncertainties are estimated using the bootstrap method with 1000 samples. In all fits, the covariance matrix is taken into account to properly handle the correlations among the data.

\begin{table}[]
	\begin{tabular}{|c|c|c|ccccccc|}
\hline
\multirow{2}{*}{$N_t$} & \multirow{2}{*}{$\beta$} & \multirow{2}{*}{$T/T_c$} & \multicolumn{7}{c|}{$\Omega_I$ [MeV]}                                                                                                                                                                                         \\ \cline{4-10} 
                       &                          &                          & \multicolumn{1}{c|}{0}                                                     & \multicolumn{1}{c|}{10} & \multicolumn{1}{c|}{15} & \multicolumn{1}{c|}{20} & \multicolumn{1}{c|}{25} & \multicolumn{1}{c|}{30} & 40                 \\ \hline
64                     & 6.5                      & $\sim 0$                 & \multicolumn{1}{c|}{\begin{tabular}[c]{@{}c@{}}PBC:9\\ OBC:1\end{tabular}} & \multicolumn{1}{c|}{-}  & \multicolumn{1}{c|}{-}  & \multicolumn{1}{c|}{1}  & \multicolumn{1}{c|}{-}  & \multicolumn{1}{c|}{1}  & 2                  \\ \hline
\multirow{6}{*}{8}     & 6.1                      & 1.02                     & \multicolumn{1}{c|}{100}                                                   & \multicolumn{1}{c|}{20} & \multicolumn{1}{c|}{20} & \multicolumn{1}{c|}{20} & \multicolumn{1}{c|}{30} & \multicolumn{1}{c|}{20} & \multirow{6}{*}{-} \\ \cline{2-9}
                       & 6.12                     & 1.05                     & \multicolumn{1}{c|}{100}                                                   & \multicolumn{1}{c|}{20} & \multicolumn{1}{c|}{20} & \multicolumn{1}{c|}{20} & \multicolumn{1}{c|}{30} & \multicolumn{1}{c|}{20} &                    \\ \cline{2-9}
                       & 6.14                     & 1.09                     & \multicolumn{1}{c|}{100}                                                   & \multicolumn{1}{c|}{40} & \multicolumn{1}{c|}{20} & \multicolumn{1}{c|}{20} & \multicolumn{1}{c|}{30} & \multicolumn{1}{c|}{20} &                    \\ \cline{2-9}
                       & 6.17                     & 1.14                     & \multicolumn{1}{c|}{100}                                                   & \multicolumn{1}{c|}{20} & \multicolumn{1}{c|}{20} & \multicolumn{1}{c|}{40} & \multicolumn{1}{c|}{30} & \multicolumn{1}{c|}{20} &                    \\ \cline{2-9}
                       & 6.2                      & 1.19                     & \multicolumn{1}{c|}{100}                                                   & \multicolumn{1}{c|}{20} & \multicolumn{1}{c|}{40} & \multicolumn{1}{c|}{40} & \multicolumn{1}{c|}{32} & \multicolumn{1}{c|}{40} &                    \\ \cline{2-9}
                       & 6.3                      & 1.38                     & \multicolumn{1}{c|}{100}                                                   & \multicolumn{1}{c|}{20} & \multicolumn{1}{c|}{40} & \multicolumn{1}{c|}{10} & \multicolumn{1}{c|}{24} & \multicolumn{1}{c|}{40} &                    \\ \hline
\end{tabular}
	\caption{Lattice setup and statistics used in this paper. The numbers of configurations are given in units of 1000. Gauge field configurations were generated using a heat bath algorithm combined with two over-relaxation steps. Neighboring saved configurations are separated by 20 update sweeps to reduce autocorrelation effects.}
	\label{table:nconf}
\end{table}

\section{Results at zero temperature}
\label{sec:results_zeroT}
\begin{figure}[!hpt]
		\includegraphics[width=0.35\textwidth]{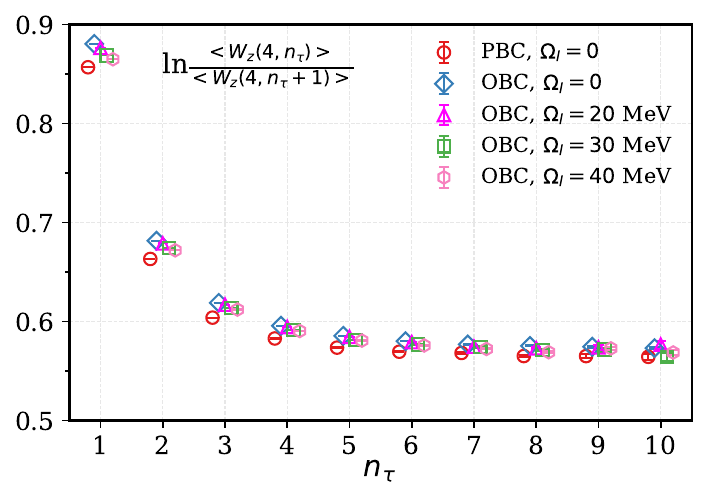}\\
		\includegraphics[width=0.35\textwidth]{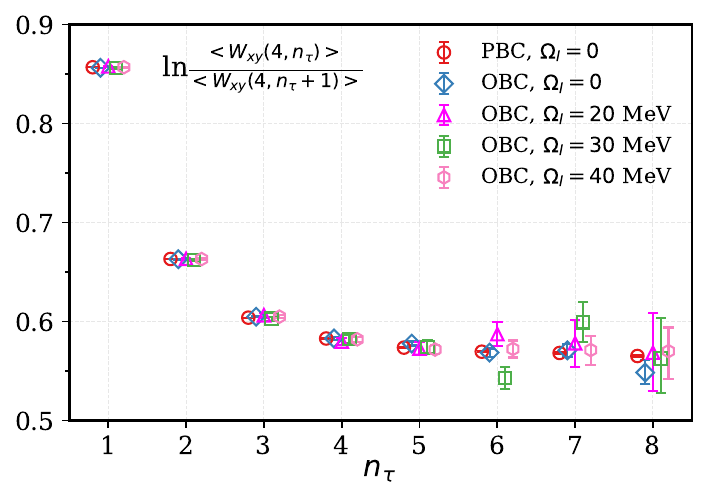}\\
		\includegraphics[width=0.35\textwidth]{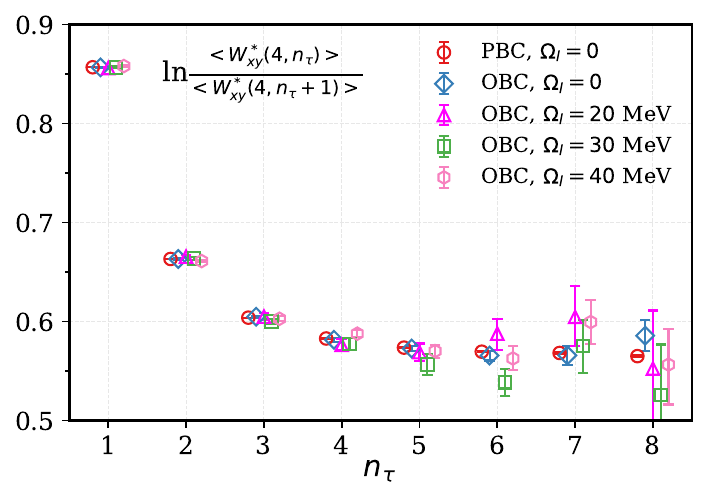}
	\caption{Zero-temperature log-ratio estimator of the static potential,
$aV^{\rm eff}_s(n;n_\tau)=\ln\!\left[\langle W_s(n,n_\tau)\rangle/\langle
W_s(n,n_\tau+1)\rangle\right]$, shown at a representative separation $n=4$ as a
function of the temporal extent $n_\tau$ for the three Wilson-loop geometries
$W_z$, $W_{xy}$, and $W_{xy}^\ast$ defined in Eqs.~\eqref{eq:Wz}--\eqref{eq:Wxy_c}
(see also~\autoref{fig:potential}). Results are shown for
$\Omega_I=0,\,20,\,30,$ and $40~\mathrm{MeV}$; the $\Omega_I=0$ PBC data on the same
lattice are included for comparison. The points are shifted slightly for visibility.}
	\label{fig:Vs_n}
\end{figure}

\begin{figure}[!htp]   
	\includegraphics[width=0.35\textwidth]{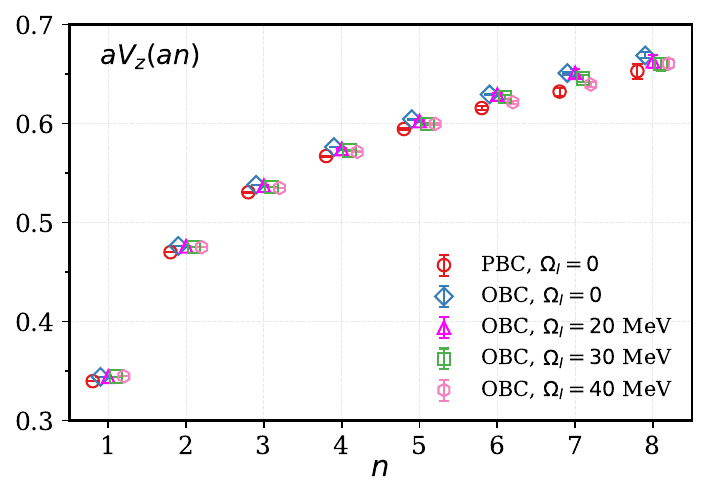}
	\includegraphics[width=0.35\textwidth]{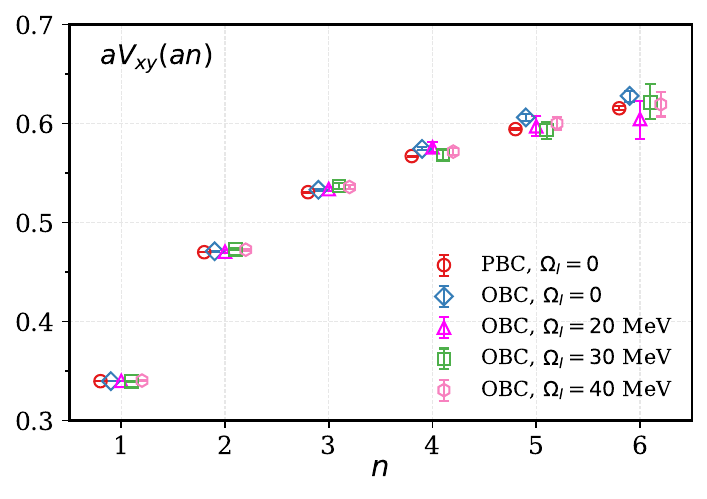}
	\includegraphics[width=0.35\textwidth]{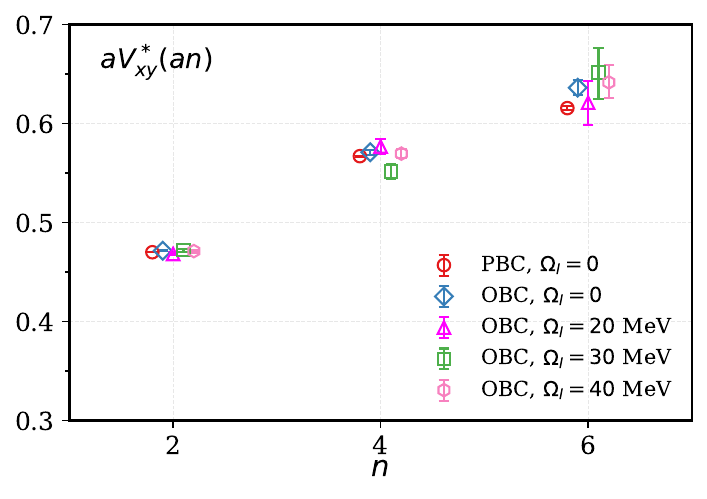}
	\caption{Zero-temperature static potential $aV_s(an)$ as a function of the
separation $n$, extracted from constant fits to the plateaus of
$aV^{\rm eff}_s(n;n_\tau)$ (cf.\ \autoref{fig:Vs_n}) for the three Wilson-loop
geometries $s=z,xy,xy^\ast$. Results are shown for
$\Omega_I=0,\,20,\,30,$ and $40~\mathrm{MeV}$; the $\Omega_I=0$ PBC result on the
same lattice is included for comparison. Within uncertainties, both the angular
velocity dependence and the differences among the three geometries are small in
the distance range studied. The points are shifted slightly for visibility.
}
	\label{fig:Vs}
\end{figure}

We first analyze the $T\simeq 0$ ensembles ($N_\tau=64$) and extract the static
quark--antiquark potential from Wilson loops. We consider the three geometries
defined in Eqs.~\eqref{eq:Wz}--\eqref{eq:Wxy_c} and illustrated in~\autoref{fig:potential} (left, middle, right): the separation parallel to the
rotation axis ($W_z$), an on-axis/off-axis pair in the transverse plane ($W_{xy}$),
and a symmetric transverse pair across the rotation axis ($W_{xy}^\ast$).

For each geometry, we form the standard log-ratio estimator,
\begin{equation}
\label{eq:Veff}
aV^{\rm eff}_s(n;n_\tau)\equiv
\ln\frac{\langle W_s(n,n_\tau)\rangle}{\langle W_s(n,n_\tau+1)\rangle},
\qquad s=z,xy,xy^\ast,
\end{equation}
which approaches $aV_s(an)$ in the large-$n_\tau$ limit.

\autoref{fig:Vs_n} shows $aV^{\rm eff}_s$ at a representative distance $n=4$
as a function of $n_\tau$ for $\Omega_I=0,\,20,\,30,$ and $40~\mathrm{MeV}$ in each of the three loop geometries. In all cases a stable plateau emerges at
sufficiently large $n_\tau$, indicating that excited-state contamination is
under control in the plateau window used below. We also include the $\Omega_I=0$
result obtained with periodic boundary conditions (PBC) on the same lattice size
to gauge possible boundary artifacts from the open boundary conditions (OBC) in $x$ and $y$ directions.

For each separation $n$ we determine $aV_s(an)$ by performing a constant fit to $aV^{\rm eff}_s(n;n_\tau)$ over the plateau
region. The resulting static potentials are shown in~\autoref{fig:Vs} as functions of $n$ for $\Omega_I=0,20,30$, and $40~\mathrm{MeV}$, together with the
$\Omega_I=0$ PBC result for comparison. 
Within the statistical precision and the range of separations studied, we
resolve no significant dependence of the static potential on the imaginary
angular velocity up to $\Omega_I=40~\mathrm{MeV}$.
The potentials extracted from the three geometries, $W_z$, $W_{xy}$, and
$W_{xy}^\ast$, also agree within errors. Thus, any rotation-induced anisotropy of the $T\simeq0$ static potential is
below our present sensitivity. Finally, the agreement between the OBC and PBC results at $\Omega_I=0$
indicates that open-boundary effects are mild for the zero-temperature
Wilson-loop observables and source separations considered here.

We note that the observed insensitivity of the $T\simeq 0$ static potential to $\Omega_I$ is consistent with the expectation that the cold vacuum does not rotate~\cite{Chernodub:2016kxh}. In the zero-temperature limit, there is no thermal medium that can be set into rotation. Our results are compatible with this physical picture, showing no statistically resolved modification of the confining interaction over the range of angular velocities and source separations studied.

\section{Results at Finite Temperature}
\label{sec:results_finiteT}

We now turn to the finite-temperature ensembles ($N_\tau=8$) and extract the
color-averaged quark--antiquark free energies from Polyakov-loop correlators,
using the three geometries introduced in Sec.~\ref{subsec:finiteT_conf} and shown in~\autoref{fig:potential}. 
Unlike the $T\simeq0$ Wilson-loop potentials, these finite-temperature
observables are spatially resolved in the transverse plane. Since the open boundaries in the $x$ and $y$ directions break transverse translational
invariance, they induce an $R_{xy}$ dependence even at $\Omega_I=0$. We therefore use axial--diagonal comparisons and bulk-region fits below to
separate the rotation-induced response from boundary effects.

We present unrenormalized free energies throughout. This is sufficient for our
purposes for two reasons. (i) At fixed temperature, all correlators share the
same additive renormalization constant, so relative comparisons at the same $T$
are meaningful. (ii) For rotation-induced changes we work with differences
between $\Omega_I\neq 0$ and $\Omega_I=0$, for which the additive constant cancels.
In what follows we assume that this additive renormalization depends only on $T$
(and not on $\Omega_I$ or the transverse position); this assumption can be tested
a posteriori by the consistency of the subtracted observables in the bulk region.

\subsection{$F_z(R_{xy},r_z)$: boundary diagnostics and bulk fits}
\label{subsec:Fz_finite}
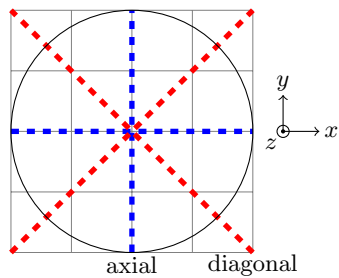
\begin{figure}[!htp]
\centering
\begin{tikzpicture}[scale=0.8]
\draw[step=1,help lines] (0,0) grid (4,4);
\draw[blue,line width=2pt,dashed] (0,2)--(4,2);
\draw[blue,line width=2pt,dashed] (2,0)--(2,4);
\draw[red,line width=2pt,dashed] (0,0)--(4,4);
\draw[red,line width=2pt,dashed] (4,0)--(0,4);
\draw (2,2) circle (2);
\node at (2,-0.2) {axial};
\node at (4,-0.2) {diagonal};
\draw[-to] (4.5,2)--(4.5,2.6);
\node at (4.5,2.8){$y$};
\draw[-to] (4.5,2)--(5.1,2);
\node at (5.3,2){$x$};
\draw (4.5,2)circle(0.1);
\fill[black](4.5,2)circle(1pt);
\node at(4.3,1.8){$z$};
\end{tikzpicture}
\caption{Schematic of the two primary directions defined in \autoref{eq:vd}. The dashed blue lines (axial) denote paths along the lattice axes, while the dashed red lines (diagonal) bisect the lattice coordinates. The black circle illustrates that the axial directions encounter the lattice boundaries at a shorter distance than the diagonal directions. Consequently, finite-volume (boundary) effects manifest at a smaller physical correlation length along the axes than along the diagonals.}
\label{fig:vd}
\end{figure}

We begin with the longitudinal geometry $F_z(R_{xy},r_z)$,~\autoref{eq:Fz_def}.
With open boundaries in $x$ and $y$, translational invariance in the transverse plane is
broken, and $F_z(R_{xy},r_z)$ acquires an explicit dependence on $R_{xy}$ already at $\Omega_I=0$.
This baseline $R_{xy}$ dependence originates from boundary effects and must be disentangled from the rotation-induced response in the analysis below.

To identify the onset of boundary artifacts, we compare data with the same coordinate distance $R$ from the rotation axis along two directions in the $x$–$y$ plane:
transverse distance from the origin along two directions in the $x$--$y$ plane:
\begin{equation}
\begin{split}
{\rm axial:}\quad &(0,0)\ \rightarrow\ (0,\pm R),\,(\pm R,0),\\
{\rm diagonal:}\quad &(0,0)\ \rightarrow\ (\pm R,\pm R),
\end{split}
\label{eq:vd}
\end{equation}
as illustrated in \autoref{fig:vd}. 
Since diagonal directions probe regions that are farther from the transverse boundaries than axis-aligned directions at the same nominal distance $R$, boundary effects are expected to set in earlier along the axial directions. The comparison between the two therefore provides a practical
criterion for identifying a bulk region with reduced boundary contamination.

The two-variable dependence of $F_z(R_{xy},r_z)$ complicates a compact characterization
of the rotation effect. For $T>T_c$, we therefore focus on the $r_z\to\infty$ limit, which
is governed by the single-Polyakov-loop expectation value at fixed $R_{xy}$:
\begin{equation}
F_z(R_{xy},\infty)=-2T\ln\left\langle \frac{1}{N_z}\sum_{z=0}^{N_z-1}\tilde{L}(R_{xy},z)\right\rangle .
\label{eq:Fz_inf}
\end{equation}

For the subtracted quantities in this section we use the convention
\begin{equation}
\Delta F_z(R_{xy}) \equiv
F_z(R_{xy},\infty)\big|_{\Omega_I=0}
-
F_z(R_{xy},\infty)\big|_{\Omega_I\neq 0}.
\label{eq:dFz_def}
\end{equation}
With this convention, positive values of $\Delta F_z$ correspond to a suppression of
the longitudinal free energy by imaginary rotation. This subtraction cancels the
additive renormalization at fixed $T$, reduces sensitivity to the full two-point
geometry, and isolates the rotation-induced response at fixed transverse position.
Since the theory is invariant under $\Omega_I\to-\Omega_I$, $\Delta F_z$ is
expected to be an even function of $\Omega_I$ in the small-$\Omega_I$ regime.

\begin{figure}[!thb]
    \includegraphics[width=0.4\textwidth]{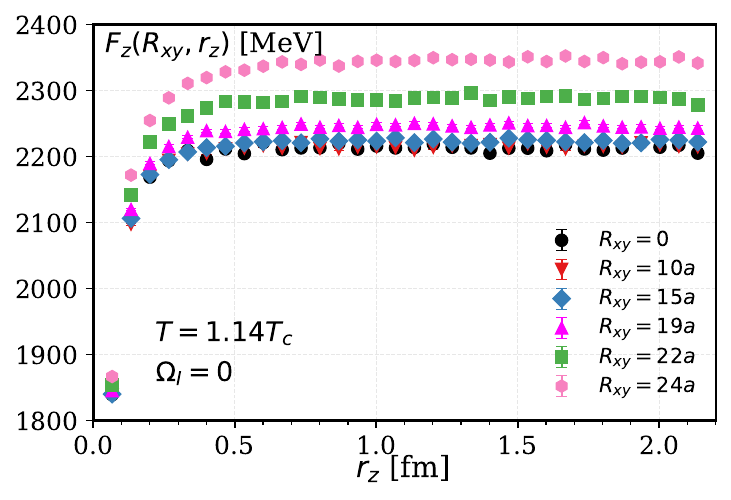} \\
    \includegraphics[width=0.4\textwidth]{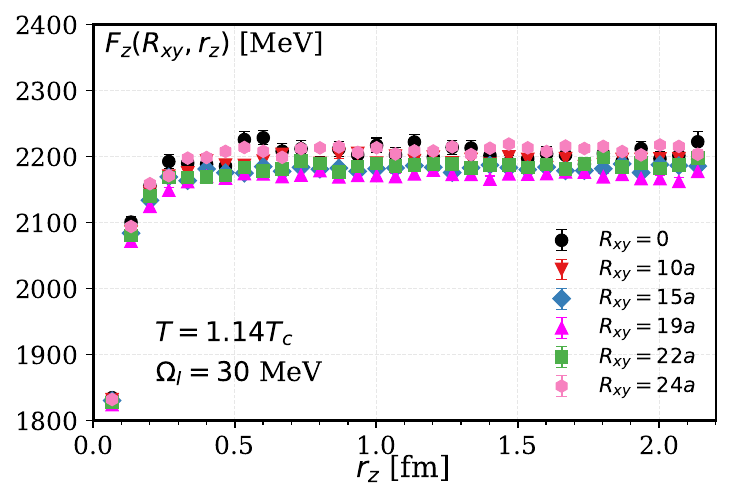}
   
    \caption{Unrenormalized longitudinal free energy $F_z(R_{xy},r_z)$ at $T=1.14T_c$ in the axial (lattice-axis) direction for several transverse positions $R_{xy}$, comparing $\Omega_I=0$ (top) and $\Omega_I=30$~MeV (bottom). At fixed $R_{xy}$, $F_z(R_{xy},r_z)$ approaches a large-$r_z$ plateau associated with color screening. The pronounced $R_{xy}$-dependence already at $\Omega_I=0$ is a boundary artifact induced by open boundaries in the transverse plane.}
    \label{fig:617Fz}
\end{figure}

\autoref{fig:617Fz} illustrates the main complication in extracting rotation effects with open boundary
conditions: even at $\Omega_I=0$, the longitudinal free energy exhibits a dependence on $R_{xy}$, and before taking the large-$r_z$ limit it depends on both $R_{xy}$ and $r_z$. For $T>T_c$, the Polyakov-loop correlator is screened, and $F_z(R_{xy},r_z)$ rapidly approaches
a constant at large $r_z$. In this regime we characterize the rotation dependence through the
large-$r_z$ limit $F_z(R_{xy},\infty)$, or equivalently through the single-Polyakov-loop contribution in Eq.~(\ref{eq:Fz_inf}).

\begin{figure}[t]
    \includegraphics[width=0.4\textwidth]{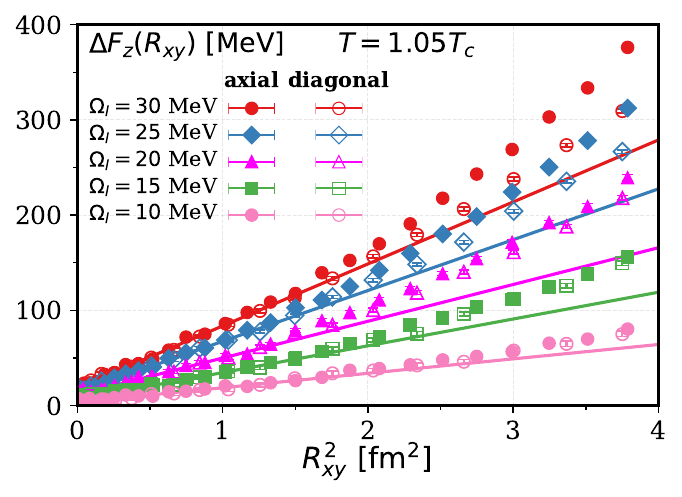}\\
    \includegraphics[width=0.4\textwidth]{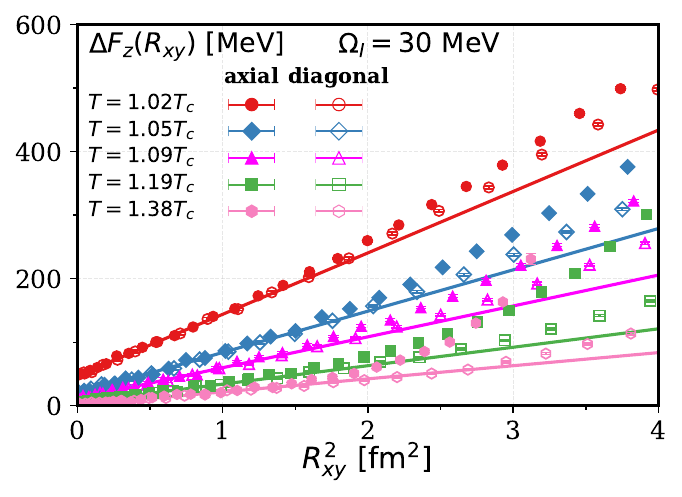}
    \caption{$\Delta F_z(R_{xy})$ defined in Eq.~\eqref{eq:dFz_def}. Top: fixed $T=1.05T_c$ and varying $\Omega_I$.
    Bottom: fixed $\Omega_I=30$~MeV and varying $T$.
    Filled (axial) points are taken along lattice axes, while open (diagonal) points are taken along lattice diagonals.
    Curves show fits to $\Delta F_z(R_{xy})=A R_{xy}^2+B$ performed in a bulk window where axial and diagonal data
    are mutually consistent.}
    \label{fig:61dFz_dFz30}
\end{figure}

The axial--diagonal comparison in \autoref{fig:61dFz_dFz30} provides a data-driven diagnostic for boundary
contamination. Since diagonal rays
remain farther from the transverse boundaries than axis-aligned rays at the same nominal radial distance
(cf.\ \autoref{fig:vd}), boundary effects are expected to set in earlier along the axes. Indeed, at small $R_{xy}$ the axial and diagonal data are mutually consistent and exhibit
approximately $\Omega_I^2$ scaling, while at larger
$R_{xy}$ the axial data depart from this behavior before the diagonal data. We interpret this loss of
axial--diagonal consistency as the onset of sizable boundary contamination, and restrict the quantitative fits to the
corresponding bulk region.

\begin{figure}[t]
    \includegraphics[width=0.4\textwidth]{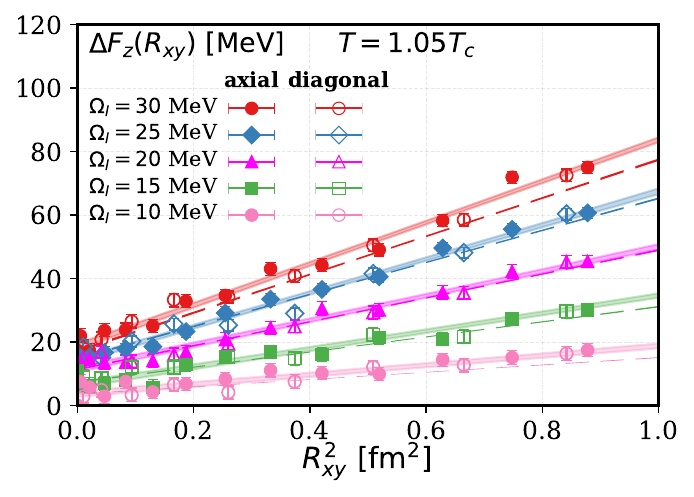}\\
    \includegraphics[width=0.4\textwidth]{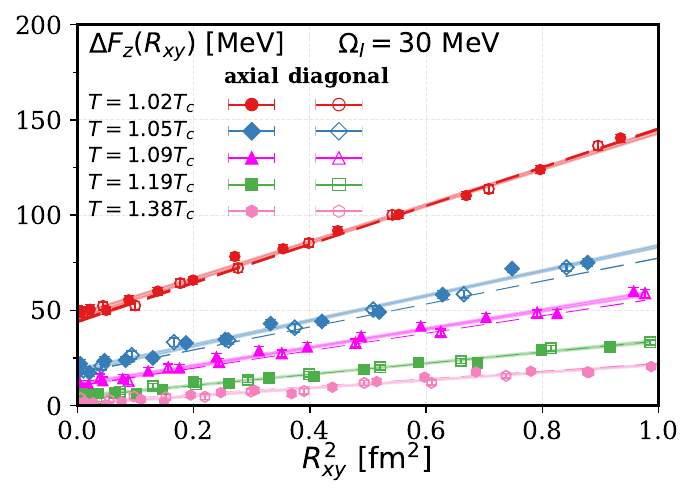}
    \caption{Same as \autoref{fig:61dFz_dFz30} but in a smaller $R_{xy}$-range. The full bands are the fit results using Eq.~\eqref{eq:dFz_quad} for each $(T,~\Omega_I)$ while the dashed bands are the global fit results using Eq.~\eqref{eq:ansatz_global}.}
    \label{fig:61dFz_dFz30_magnified}
\end{figure}

In the bulk region, we find that $\Delta F_z(R_{xy})$ is described within
errors by the quadratic form
\begin{equation}
\Delta F_z(R_{xy}) = A\,R_{xy}^2 + B\,,
\label{eq:dFz_quad}
\end{equation}
where the coefficients $A$ and $B$ depend on $T$ and $\Omega_I$. 
Therefore, for each $(T,\Omega_I)$, we fit the lattice data to Eq.~\eqref{eq:dFz_quad} over the largest admissible $R_{xy}$ window satisfying the axial--diagonal consistency criterion. The fits are performed using a bootstrap analysis with 1000 samples, accounting for correlations among different $R_{xy}$ values. The resulting fits are shown as colored bands in~\autoref{fig:61dFz_dFz30} and also in~\autoref{fig:61dFz_dFz30_magnified} for better visibility in a narrow region of $R_{xy}^2$. We have verified that further reducing the fit window leaves the extracted results stable, with only negligible variations. The individual bulk-window fits yield an average $\chi^2/{\rm dof}$ of 1.98.

\begin{figure}[!thb]
    \includegraphics[width=0.4\textwidth]{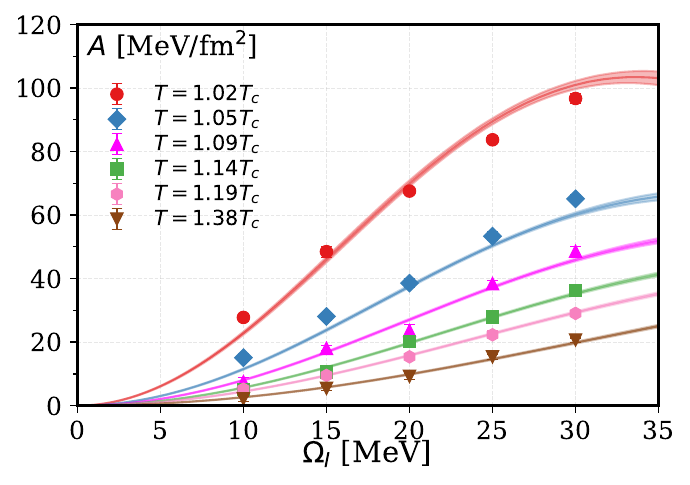}\\
    \includegraphics[width=0.4\textwidth]{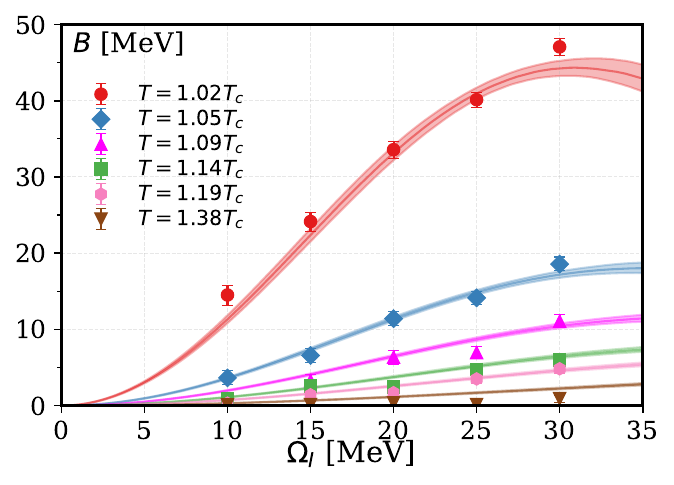}
    \caption{Fit parameters $A$ and $B$ from Eq.~\eqref{eq:dFz_quad} as functions of $\Omega_I$ for several temperatures.
    The data points are individual fit results of Eq.~\eqref{eq:dFz_quad} for each $(T,\Omega_I)$. The fit bands indicate the global fit of Eq.~\eqref{eq:ansatz_global}.}
    \label{fig:AB_Omega}
\end{figure}

\begin{figure}[!thb]
    \includegraphics[width=0.4\textwidth]{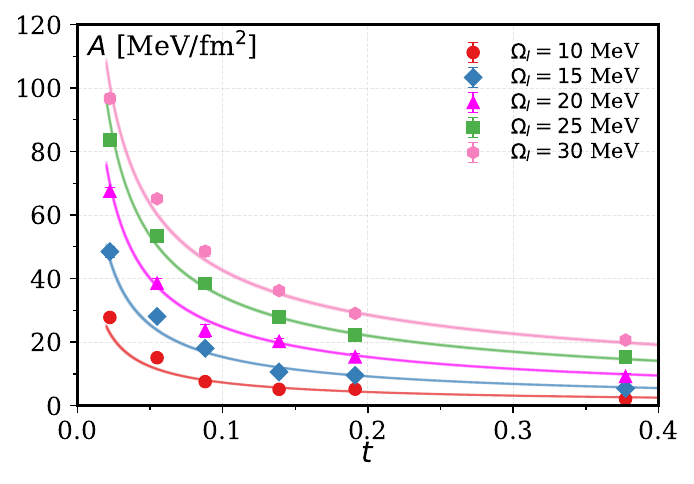}\\
    \includegraphics[width=0.4\textwidth]{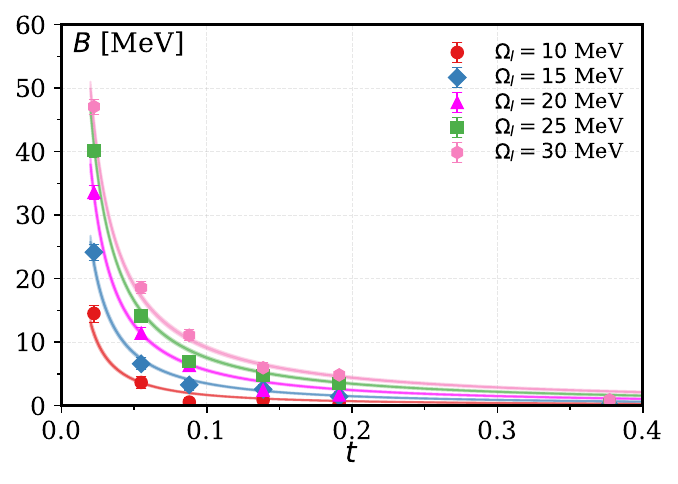}
    \caption{Fit parameters $A$ and $B$ from Eq.~\eqref{eq:dFz_quad} as functions of reduced temperature
    $t=(T-T_c)/T_c$ for several $\Omega_I$. The data points are individual fit results of Eq.~\eqref{eq:dFz_quad} for each $(T,\Omega_I)$. The fit bands indicate the global fit of Eq.~\eqref{eq:ansatz_global}.}
    \label{fig:AB_beta}
\end{figure}

\autoref{fig:AB_Omega} and \ref{fig:AB_beta} show the obtained fit
parameters $A$ and $B$ as functions of $\Omega_I$ and of the reduced
temperature $t=(T-T_c)/T_c$ via Eq.~\eqref{eq:dFz_quad}. The fit results of $A$ and $B$, shown as data points, exhibit two robust trends. First, at fixed temperature both
parameters increase with the magnitude of the imaginary angular
velocity. Since the theory is invariant under $\Omega_I\to-\Omega_I$,
the rotation-induced correction to scalar observables such as the
Polyakov-loop free energy is expected to be an even function of
$\Omega_I$ in the small-$\Omega_I$ regime. Second, as shown more
directly in \autoref{fig:AB_beta}, at fixed $\Omega_I$ both $A$ and
$B$ decrease as the temperature is raised above $T_c$. This indicates
that the rotation-induced modification of the free energy is maximal near the transition and becomes weaker deeper in the deconfined phase.
The decrease of $B$ with increasing $t$ is  faster than that of
$A$, suggesting that the approximately $R_{xy}$-independent component
of the response is more strongly concentrated near $T_c$.

The leading structure of the fit can be motivated by the
rotating-frame formulation. Rotation enters the lattice action through
the external metric associated with the rotating coordinate system. In
the bulk region and for sufficiently small angular velocity, the
leading local scalar constructed from the angular velocity and the distance
to the rotation axis is proportional to the square of the local velocity scale,
$(\Omega R_{xy})^2$. Together with the symmetry under
$\Omega\to-\Omega$, this suggests the leading form
\begin{equation}
    \Delta F_z(R_{xy})
    =
    \Omega_I^2\left[c_2(T)R_{xy}^2+c_0(T)\right]
    +O(\Omega_I^4),
\end{equation}
where $c_2(T)$ and $c_0(T)$ encode the thermal response of the
Polyakov-loop free energy to the rotating background. This provides a
natural parametrization for the bulk behavior, with
$\Delta F_z(R_{xy})=AR_{xy}^2+B$ and $A$ and $B$ depending on
$T$ and $\Omega_I$.

Motivated by the observed enhancement of the response when approaching
$T_c$ from above, we parametrize the temperature dependence of $A$ and
$B$ by effective power laws,
\begin{equation}
\begin{split}
    A(t,\Omega_I)&=A'(\Omega_I)t^{-a(\Omega_I)},\\
    B(t,\Omega_I)&=B'(\Omega_I)t^{-b(\Omega_I)}.
\end{split}
\label{AB_ansatz}
\end{equation}
These exponents should not be interpreted as universal critical exponents.
Rather, they provide a compact description of the near-transition
enhancement over the temperature range covered by the simulations.

Guided by the evenness under $\Omega_I\to-\Omega_I$, we further parametrize the $\Omega_I$ dependence as
\begin{equation}
\begin{split}
    A'(\Omega_I)&=A_2\Omega_I^2,\\
    B'(\Omega_I)&=B_2\Omega_I^2,\\
    a(\Omega_I)&=a_0-a_2\Omega_I^2,\\
    b(\Omega_I)&=b_0-b_2\Omega_I^2.
\end{split}
\label{AB_coeff_ansatz}
\end{equation}
Combining Eqs.~\eqref{AB_ansatz} and \eqref{AB_coeff_ansatz}, we obtain a parametrization
\begin{equation}
\begin{split}
    \Delta F_z(t,\Omega_I;R_{xy})
    =&~\frac{A_2\Omega_I^2}
    {t^{a_0-a_2\Omega_I^2}} R_{xy}^2
    +\frac{B_2\Omega_I^2}
    {t^{b_0-b_2\Omega_I^2}}, 
\end{split}
\label{eq:ansatz_global}
\end{equation}
that allows us to perform a global fit using the data obtained at all the values of $T$ and $\Omega_I$. We combine all the bootstrap samples, which were used in the separate fits using Eq.~\eqref{eq:dFz_quad} at each $T$ and $\Omega_I$, to perform the global fit using Eq.~\eqref{eq:ansatz_global} in the same $R_{xy}$-range. We have used 896 different $(t,\Omega_I;R_{xy})$ data points to fit 6 parameters $A_2$, $a_0$, $a_2$, $B_2$, $b_0$, $b_2$. The global fit results, which are shown as solid curves in \autoref{fig:AB_Omega} and \ref{fig:AB_beta}, show that this
symmetry-guided parametrization captures the main trends of the obtained $A$ and $B$ obtained from fits using Eq.~\eqref{eq:dFz_quad} over the
explored range of temperatures and imaginary angular velocities. The consistency of the fits using Eq.~\eqref{eq:dFz_quad} for each ($T,\Omega_I$) and the global fits using Eq.~\eqref{eq:ansatz_global} can also be observed in the $R_{xy}^2$ dependence of $\Delta F_z(R_{xy})$ shown in ~\autoref{fig:61dFz_dFz30_magnified}.
The
resulting global fit parameters are listed in Table~\ref{table1}. All fitted
coefficients are positive in this range. Given the value of
$\chi^2/\mathrm{dof}$, however, Eq.~\eqref{eq:ansatz_global} should be viewed as
a compact phenomenological interpolation of the observed behavior, rather than as a
precision fit or a controlled scaling form.

The term proportional to $R_{xy}^2$ in Eq.~\eqref{eq:dFz_quad} has a
simple qualitative interpretation. In the bulk region, the
rotation-induced shift grows with the distance from the rotation axis,
consistent with a leading dependence on the local velocity scale; in the
imaginary-rotation simulations this scale is represented by
$(\Omega_I R_{xy})^2$. This behavior should not be interpreted as the
classical rotational energy of a static quark--antiquark pair. Our calculation is formulated in the co-rotating frame, where the thermal
medium is stationary and rotation enters through the nontrivial Euclidean metric. The static free energy is therefore defined
with respect to the Euclidean evolution in this non-inertial frame,
rather than the Hamiltonian of a static medium in an
inertial frame. We thus interpret the fitted $R_{xy}^2$ term as the leading spatial response to the rotating background in the
range of $R_{xy}$ and $\Omega_I$ explored here.

The intercept $B$ parametrizes the part of $\Delta F_z$ that is
approximately independent of $R_{xy}$ within the selected bulk window.
It may contain a genuine $R_{xy}$-independent contribution to the
Polyakov-loop free energy in the rotating background. However, because
the free energies used here are not renormalized by short-distance
matching, $B$ may also absorb residual renormalization effects, including
a possible $\Omega_I$ dependence of the additive constant. We therefore
retain $B$ as a fit parameter, but do not assign it a separate
physical interpretation.

Finally, we comment on the relation to real angular velocity. 
Since the
simulations are performed at imaginary angular velocity, any
continuation to real rotation should be based on a Taylor expansion
around $\Omega=0$. At fixed $T$ and $R_{xy}$, the symmetry under
$\Omega_I\to-\Omega_I$ implies
\begin{equation}
    \Delta F_z(T,\Omega_I;R_{xy})
    =
    C_2(T,R_{xy})\Omega_I^2
    +
    C_4(T,R_{xy})\Omega_I^4
    +
    O(\Omega_I^6).
\end{equation}
Analytic continuation to real angular velocity is then implemented by
$\Omega_I^2\to -\Omega^2$.  With our convention
$\Delta F_z=F_z|_{\Omega_I=0}-F_z|_{\Omega_I\neq0}$, this gives
\begin{equation}
    \Delta F_z^{\rm real}(T,\Omega;R_{xy})
    =
    -C_2(T,R_{xy})\Omega^2
    +
    C_4(T,R_{xy})\Omega^4
    +
    O(\Omega^6).
\end{equation}
Equivalently, the shift of the free energy at real angular velocity is
\begin{equation}
\begin{aligned}
    &F_z^{\rm real}(T,\Omega;R_{xy})-F_z(T,0;R_{xy})  \\
    &\qquad =
    C_2(T,R_{xy})\Omega^2
    - C_4(T,R_{xy})\Omega^4
    + O(\Omega^6).
\end{aligned}
\end{equation}
Therefore, a positive leading Taylor coefficient $C_2$ extracted from
the imaginary-rotation data would imply an enhancement of $F_z$ at
leading order for real rotation, provided analyticity around
$\Omega=0$ holds and the quadratic term dominates.

We stress that Eq.~\eqref{eq:ansatz_global} is introduced only as a compact
interpolation of the imaginary-$\Omega_I$ data over the explored range. Although it is an even function
of $\Omega_I$ and can be formally continued via $\Omega_I^2\!\to\!-\Omega^2$ for
$T>T_c$, its $\Omega_I$-dependent exponents are not derived from a controlled
small-$\Omega$ expansion. As a result, quantitative predictions at real
angular velocity are necessarily ansatz-dependent, especially in the vicinity
of $T_c$. A controlled analytic continuation would require a dedicated
small-$\Omega_I$ determination of the Taylor coefficients $C_{2n}(T,R_{xy})$, including systematic studies of truncation and fit-range
dependence, which is beyond
the scope of the present work. The robust conclusion from
Figs.~\ref{fig:AB_Omega} and \ref{fig:AB_beta} is that, for $T>T_c$, imaginary
rotation suppresses $F_z$, and this suppression weakens as the temperature
increases.

\begin{table}[!htp]
\begin{tabular}{|c|c|}
\hline
$A_2$ [MeV]        & 489(11)                   \\ \hline
$a_0$                  & 0.787(10)                 \\ \hline
$a_2$ [MeV$^{-2}$] & 2.336(91)$\times 10^{-4}$ \\ \hline
$B_2$ [MeV$^{-1}$] & 8.92(70)$\times 10^{-4}$  \\ \hline
$b_0$                  & 1.303(27)                 \\ \hline
$b_2$ [MeV$^{-2}$] & 2.74(18)$\times 10^{-4}$  \\ \hline
$\chi^2$/dof    & 2.35(10)                  \\ \hline
\end{tabular}
\caption{Fit parameters for $\Delta F_z(t,\Omega_I;R_{xy})$ using the ansatz in Eq.~\ref{eq:ansatz_global}.}
\label{table1}
\end{table}

\subsection{$F_{xy}(r_{xy})$ and $F_{xy}^*(r_{xy})$}
\begin{figure*}
    \centerline{
    \includegraphics[width=0.33\textwidth]{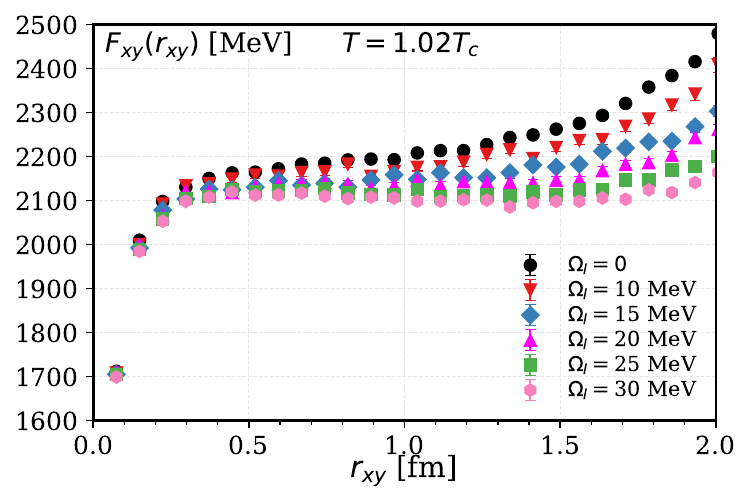}
    \includegraphics[width=0.33\textwidth]{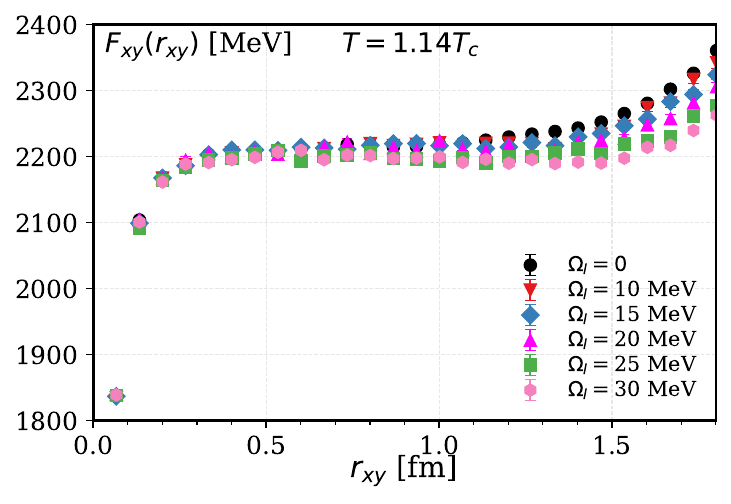}
    \includegraphics[width=0.33\textwidth]{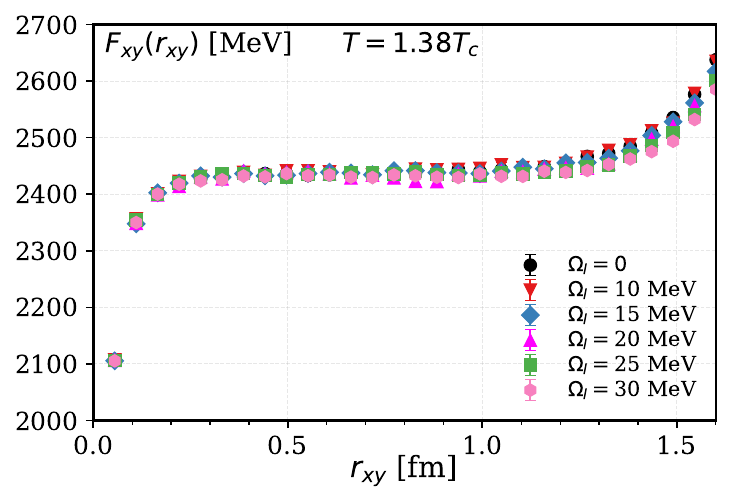}
    }
    \caption{Unrenormalized transverse free energy $F_{xy}(r_{xy})$ at
    $T=1.02T_c$, $1.14T_c$, and $1.38T_c$ (from left to right) for several values of
    $\Omega_I$. The data are taken along the axial direction in the
    transverse plane. Imaginary rotation suppresses $F_{xy}$, and the
    effect becomes weaker as the temperature is increased.
    }
    \label{fig:Fxy}
\end{figure*}

We now turn to the transverse geometries, $F_{xy}(r_{xy})$ and
$F_{xy}^*(r_{xy})$, defined in Sec.~\ref{subsec:finiteT_conf}. In contrast to
$F_z(R_{xy},r_z)$, where the quark--antiquark separation is along the
rotation axis and the radial distance $R_{xy}$ can be varied
independently, the transverse free energies involve separations in the
plane perpendicular to the rotation axis. As a result, varying
$r_{xy}$ generally changes both the $Q\bar Q$ separation and the
radial locations of the static sources with respect to the rotation
axis. This geometric mixing makes the transverse channels less suitable
for a direct quantitative fit analogous to Eq.~\eqref{eq:dFz_quad}, but they provide an important
independent cross-check of the rotation effects observed in the longitudinal
channel.

\autoref{fig:Fxy} shows $F_{xy}(r_{xy})$ at
$T=1.02T_c$, $1.14T_c$, and $1.38T_c$. As in the longitudinal channel,
imaginary rotation lowers the bare free energy in the range of parameters studied. The suppression is most
pronounced close to $T_c$ and becomes weaker as the temperature is
increased. This trend is consistent with the temperature
dependence of the fit parameters $A$ and $B$ obtained from 
$\Delta F_z(R_{xy})$.

To quantify the rotation-induced change in the transverse channels, we
define
\begin{equation}
    \Delta F_{xy}(r_{xy})
    =
    F_{xy}(r_{xy})\big|_{\Omega_I=0}
    -
    F_{xy}(r_{xy})\big|_{\Omega_I\neq0},
    \label{eq:dFxy_def}
\end{equation}
and analogously
\begin{equation}
    \Delta F_{xy}^*(r_{xy})
    =
    F_{xy}^*(r_{xy})\big|_{\Omega_I=0}
    -
    F_{xy}^*(r_{xy})\big|_{\Omega_I\neq0}.
    \label{eq:dFxy_star_def}
\end{equation}
With this convention, positive values of $\Delta F_{xy}$ and
$\Delta F_{xy}^*$ correspond to a lowering of the corresponding bare transverse free
energies induced by imaginary rotation.

\begin{figure}[!htb]
    \includegraphics[width=0.4\textwidth]{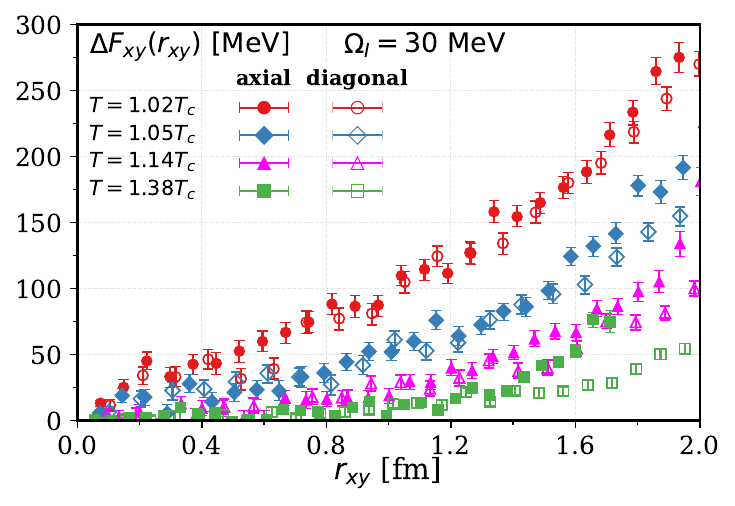}\\
    \includegraphics[width=0.4\textwidth]{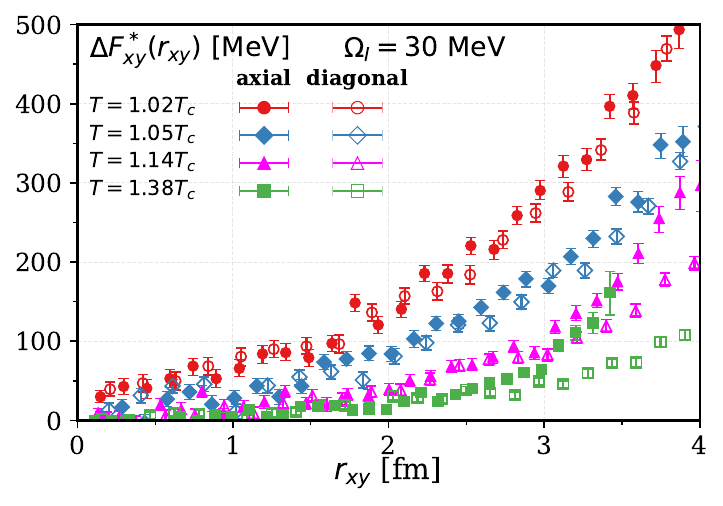}
    \caption{$\Delta F_{xy}(r_{xy})$ (top) and $\Delta F_{xy}^*(r_{xy})$ (bottom) at
    $\Omega_I=30$ MeV for several temperatures. A positive value denotes
    a suppression of the transverse free energy by imaginary rotation.
    In the deconfined phase, the magnitude of the suppression decreases
    as the temperature is increased.}
    \label{fig:Fxy_beta}
\end{figure}

The two transverse observables correspond to different source geometries. In
$F_{xy}(r_{xy})$, one static source lies on the rotation axis,
while the other is displaced by $r_{xy}$. Thus increasing $r_{xy}$
changes both the pair separation and the distance of one source from the rotation axis. In $F_{xy}^*(r_{xy})$, the two sources are placed symmetrically about the rotation axis; the pair separation is $r_{xy}$, and each source is located at a distance $r_{xy}/2$ from the axis. Therefore, at fixed $r_{xy}$, comparing $\Delta F_{xy}$ and $\Delta F_{xy}^*$ probes the sensitivity of the
rotation-induced shift to the radial arrangement of the sources at
fixed pair separation. This comparison does not by itself isolate a
single-source contribution; for that purpose, the longitudinal shift $\Delta F_z(R_{xy})$ provides the cleaner reference through the factorization test below.

\begin{figure*}[!htp]
    \includegraphics[width=0.329\textwidth]{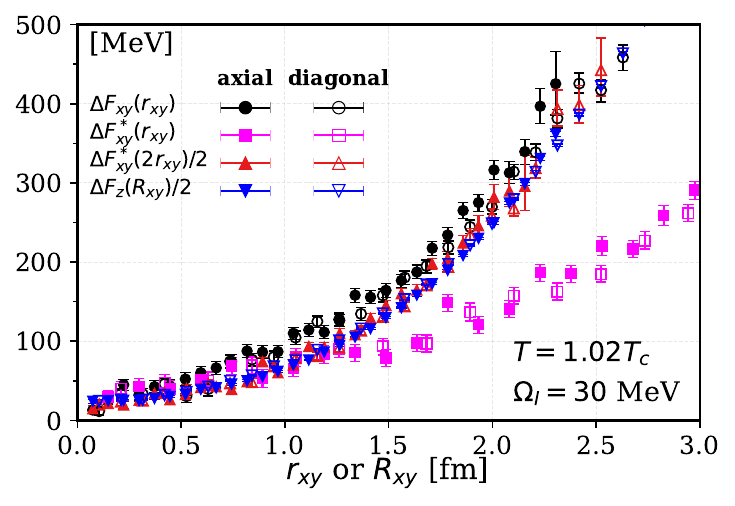}
    \includegraphics[width=0.329\textwidth]{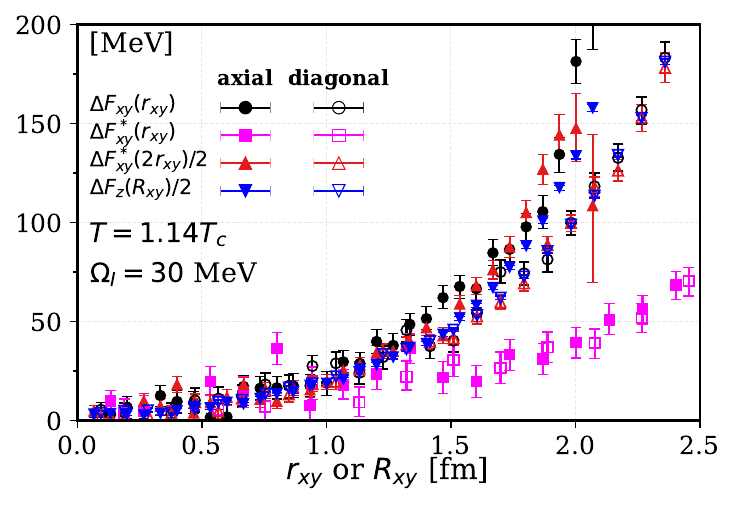}
    \includegraphics[width=0.329\textwidth]{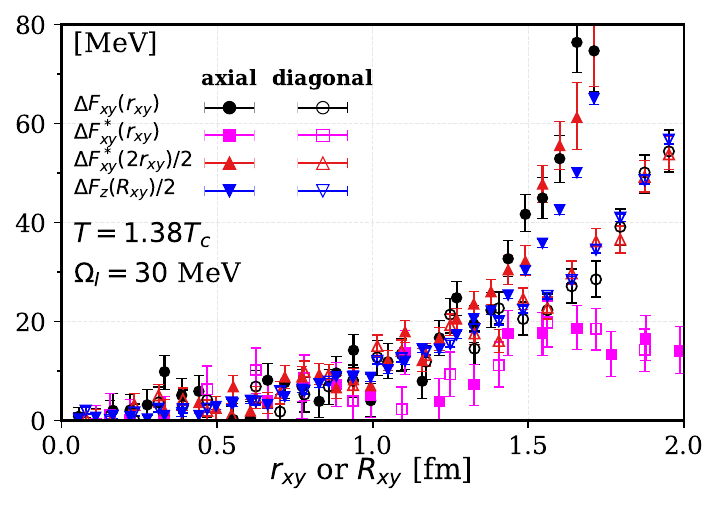}
    \caption{Comparison of $\Delta F_{xy}(r_{xy})$,
$\Delta F_{xy}^*(r_{xy})$, $\Delta F_{xy}^*(2r_{xy})/2$, and
$\Delta F_z(R_{xy}=r_{xy})/2$  at
    $\Omega_I=30$ MeV for $T=1.02T_c$, $1.14T_c$, and $1.38T_c$ (from left to right).}
    \label{fig:Fxy_Fxy_c}
\end{figure*}

Since the rotation-induced transverse shifts decrease both as $\Omega_I$ is reduced and as the temperature is raised, their signal-to-noise ratio deteriorates rapidly. We therefore present the
detailed temperature and geometry comparisons in
\autoref{fig:Fxy_beta} and \autoref{fig:Fxy_Fxy_c} at $\Omega_I=30$ MeV, where the signal is best
resolved. The $\Omega_I$ dependence of the underlying transverse free energy is already shown in \autoref{fig:Fxy}. The results at smaller $\Omega_I$ are compatible with the same
qualitative behavior but do not permit an equally discriminating comparison among the different geometries.

\autoref{fig:Fxy_beta} shows $\Delta F_{xy}(r_{xy})$ and
$\Delta F_{xy}^*(r_{xy})$ at $\Omega_I=30$ MeV for several
temperatures. Both shifts are positive wherever the effect is statistically
resolved in the range shown,
indicating a suppression
of the transverse free energies by imaginary rotation.  Moreover, the magnitude of the suppression decreases as the
temperature is raised above $T_c$. This suggests that the
weakening of rotation-induced effects at higher temperature is not
specific to the longitudinal geometry, but also appears for
quark--antiquark pairs separated in the transverse plane.

The transverse data also show an approximately increasing trend with
$r_{xy}$ in the central region where the axial and diagonal results
remain consistent. This is qualitatively consistent with the
expectation that the rotation response grows with the distance from the
rotation axis, or with the local velocity scale $\Omega_I r$. However,
unlike the longitudinal observable $\Delta F_z(R_{xy})$, the transverse
observables do not isolate a single radial coordinate. The observed
$r_{xy}$ dependence therefore contains both the effect of changing the
pair separation and the effect of moving one or both static sources
away from the rotation axis. For this reason, we do not attempt to fit
the transverse data with a direct analogue of Eq.~\eqref{eq:dFz_quad}.

Further intuition can be obtained in the deconfined phase from the
factorization of Polyakov-loop correlators at separations large compared
with the screening length. In this regime, cluster decomposition implies
\begin{equation}
\left\langle \tilde L(\mathbf{x})\tilde L^\dagger(\mathbf{y})\right\rangle
\simeq
\left\langle \tilde L(\mathbf{x})\right\rangle
\left\langle \tilde L^\dagger(\mathbf{y})\right\rangle ,
\end{equation}
up to exponentially suppressed connected contributions.
Using the large-distance definition in Eq.~\eqref{eq:Fz_inf}, we define
an effective single-quark free energy for a static source located at
radial distance $R_{xy}$ from the rotation axis,
\begin{equation}
    F_Q(R_{xy})\equiv -T\ln\langle \tilde L(R_{xy})\rangle
    =\frac{1}{2}F_z(R_{xy},\infty).
\end{equation}
Here $F_z(R_{xy},\infty)$ denotes the large-$r_z$ limit of the
longitudinal free energy at fixed transverse position $R_{xy}$.

If the transverse correlators are dominated by these single-source
contributions at large separation, the asymmetric geometry, with one source on
the rotation axis and the other a distance $r_{xy}$ away, gives
\begin{equation}
    F_{xy}(r_{xy}) \simeq F_Q(0)+F_Q(r_{xy}),
\end{equation}
while the symmetric geometry, with both sources a distance $r_{xy}/2$ from the
rotation axis, gives
\begin{equation}
    F_{xy}^*(r_{xy}) \simeq 2F_Q(r_{xy}/2).
\end{equation}
For each quantity we use the subtraction convention defined in
Eqs.~\eqref{eq:dFz_def}, \eqref{eq:dFxy_def} and \eqref{eq:dFxy_star_def}. Correspondingly, we define $\Delta F_Q(R_{xy})\equiv F_Q(R_{xy})|_{\Omega_I=0}-F_Q(R_{xy})|_{\Omega_I\ne0}$ at fixed
$R_{xy}$.  The single-source picture then gives
\begin{equation}
\begin{aligned}
    \Delta F_{xy}(r_{xy})
    &\simeq \Delta F_Q(0)+\Delta F_Q(r_{xy}) \\
    &=
    \frac{1}{2}\left[
    \Delta F_z(R_{xy}=0)
    +
    \Delta F_z(R_{xy}=r_{xy})
    \right],\\
    \Delta F_{xy}^*(2r_{xy})
    &\simeq 2\Delta F_Q(r_{xy})
     \simeq \Delta F_z(R_{xy}=r_{xy}).
\end{aligned}
\label{eq:dFxy_factorization}
\end{equation}
These relations are not used as fit formulae.  They give the expected
large-distance behavior when the connected part of the Polyakov-loop
correlator is subleading, so that the correlator is dominated by the
product of two single-Polyakov-loop expectation values.  We note,
however, that the relations in Eq.~\eqref{eq:dFxy_factorization}
concern the rotation-induced differences $\Delta F$.  Their approximate
validity does not by itself establish factorization of the full free
energy at each value of $\Omega_I$; it also allows for a residual
two-source contribution that is approximately insensitive to imaginary
rotation.

The individual transverse shifts $\Delta F_{xy}$ and
$\Delta F_{xy}^*$ were shown in \autoref{fig:Fxy_beta}.  In
\autoref{fig:Fxy_Fxy_c} we reorganize these data to test the
single-source interpretation in Eq.~\eqref{eq:dFxy_factorization}.
The most direct comparison is between $\Delta F_{xy}^*(2r_{xy})/2$ and
$\Delta F_z(R_{xy}=r_{xy})/2$.  In the former quantity the two sources
are separated by $2r_{xy}$, while each source is located at radial
distance $r_{xy}$ from the rotation axis. In the large-distance
factorization regime, both quantities estimate the same
single-source shift, $\Delta F_Q(r_{xy})$.  By contrast,
$\Delta F_{xy}(r_{xy})$ also contains the on-axis contribution
$\Delta F_Q(0)$.

As shown in \autoref{fig:Fxy_Fxy_c}, these two estimates agree
approximately over a finite range of $r_{xy}$ where the axial and
diagonal data remain mutually consistent. 
At larger $r_{xy}$, the comparison becomes less decisive with the present data.  In particular,
the axial and diagonal data sets begin to show visible differences in parts of this region, indicating increased sensitivity to the
transverse direction and boundary effects.  Since the size and onset of this axial--diagonal splitting are not uniform in temperature, we do
not identify it with a sharp boundary scale or a sharp screening length.

A natural interpretation of the single-source relation is provided by  color screening in the deconfined phase. At sufficiently large separations, the connected contribution to the Polyakov-loop correlator is expected to become subleading, and the free energy can approach a sum of two independent single-source free energies. At shorter separations, however, the quark and antiquark can remain a correlated pair, and the free energy can contain a genuine two-body contribution depending on the separation and geometry. In this case, the full free energy need not be written as a sum of two independent single-source free energies. Nevertheless, the agreement observed in Eq.~\eqref{eq:dFxy_factorization} shows that no additional rotation-induced two-source contribution is resolved within the present accuracy. At still larger separations the comparison becomes less conclusive because of increasing boundary effects and directional dependence, rather than providing evidence for a sharp breakdown of the single-source relation. 

The remaining data sets in \autoref{fig:Fxy_Fxy_c} illustrate the
geometric mixing in the transverse observables.  The quantity
$\Delta F_{xy}^*(r_{xy})$ samples sources located at radius
$r_{xy}/2$, whereas in the single-source regime
$\Delta F_{xy}^*(2r_{xy})/2$ and
$\Delta F_z(R_{xy}=r_{xy})/2$ estimate the same single-source response at
radius $r_{xy}$.  Similarly, $\Delta F_{xy}(r_{xy})$ contains the
additional on-axis contribution $\Delta F_Q(0)$ and is therefore less
direct as a test of the second relation in
Eq.~\eqref{eq:dFxy_factorization}.  Overall, at $\Omega_I=30$ MeV and over the bulk
region where the axial and diagonal data remain consistent,~\autoref{fig:Fxy_Fxy_c} supports the interpretation that, in the
deconfined bulk region, the dominant transverse response to imaginary
rotation is compatible with a radial single-source free-energy shift.

\section{Conclusion}
\label{sec:summary}

We have studied static quark--antiquark interactions in rotating
SU(3) gluodynamics using quenched lattice simulations at imaginary
angular velocity.  The calculation was performed in the rotating-frame
formulation, with open boundary conditions in the plane perpendicular
to the rotation axis.  Since these boundaries introduce transverse
inhomogeneities, we used axial--diagonal comparisons to identify a
central region in which direction-dependent boundary effects are not
statistically resolved.

At $T\simeq0$, we extracted the static potential from Wilson loops for
quark--antiquark pairs aligned with the rotation axis, for
on-axis/off-axis pairs in the transverse plane, and for symmetric
transverse pairs across the rotation axis.  Within the present
accuracy and distance range, we find no statistically significant
dependence on imaginary angular velocity up to $\Omega_I=40$ MeV or
difference among the three source geometries.  The comparison with
periodic boundary conditions at $\Omega_I=0$ further suggests that
open-boundary effects are mild for the low-temperature observables and
source separations considered here.

At finite temperature, the rotation-induced response is much more
pronounced.  In the deconfined phase, imaginary rotation suppresses the
bare color-averaged free energies obtained from Polyakov-loop
correlators.  For the longitudinal geometry, the large-distance
free-energy shift in the selected central region is well described by
a contribution proportional to $R_{xy}^2$ plus an approximately
position-independent term.  This behavior is consistent with a leading
radial dependence governed by the squared local velocity scale,
although the corresponding fit parameters remain bare response quantities at finite lattice spacing.  The transverse channels exhibit the same
qualitative suppression, while their distance dependence additionally reflects the radial arrangement of the static sources.  At $\Omega_I=30$ MeV, comparison between the symmetric transverse geometry and the longitudinal channel over the bulk range, where axial-diagonal consistency is observed, is compatible with a radial single-source free-energy shift.

A common trend emerges from all finite-temperature observables studied:
the rotation-induced modification is strongest near $T_c$ and becomes
weaker as the temperature is raised deeper into the deconfined phase.
The contrast between the low- and finite-temperature results suggests
that the observed finite-temperature effect is predominantly associated
with the response of the thermal gluonic medium.  A smaller
rotation-induced modification of the low-temperature static potential
cannot be excluded at the present precision. Because the present analysis uses bare free energies and
does not include continuum or finite-volume extrapolations, this
interpretation should remain qualitative.

Finally, we emphasize that our results are obtained at imaginary
angular velocity.  Under the usual assumption of analyticity around
$\Omega=0$, the leading $\Omega^2$ contribution at real angular
velocity would have the opposite sign from that at imaginary angular
velocity.  A quantitative analytic continuation, however, requires a
dedicated small-$\Omega_I$ determination of the Taylor coefficients
and systematic studies of truncation, fit-window, cutoff, and boundary
effects.  We therefore limit the present conclusions to the
imaginary-rotation setup.  Future work should include continuum and
finite-volume studies, renormalized Polyakov-loop observables, and
extensions to full QCD with dynamical quarks.

\section*{Acknowledgements}
We thank Defu Hou and Xu-Guang Huang for interesting discussions. This work is supported partly by the National Natural Science Foundation of China under Grants No. 12325508, No. 12293064 and No. 12293060, as well as the National Key Research and Development Program of China under Contract No. 2022YFA1604900 and the Fundamental Research Funds for the Central Universities, Central China Normal University under Grants No. 30101250314 and No. 30106250152. 
O.K. acknowledges support by the Deutsche Forschungsgemeinschaft (DFG, German Research Foundation) through the CRC-TR 211 `Strong-interaction matter under extreme conditions' – project number 315477589 – TRR 211.
The numerical simulations have been performed on the GPU cluster in the Nuclear Science Computing Center at Central China Normal University ($\mathrm{NSC}^{3}$).

\section*{Data Availability}
The data that support the findings of this article are openly available \cite{kaczmarek2026dataset}.

\bibliographystyle{JHEP}
\bibliography{HQPotential_Rotation}

\end{document}